\documentclass[a4paper,12pt]{article}
\usepackage{jheppub}
\usepackage{amsmath,amssymb,amsfonts}
\usepackage{mathrsfs}%
\usepackage{upgreek}
\usepackage{hyperref}
\hypersetup{colorlinks=true,linkcolor=magenta,anchorcolor=green,citecolor=cyan,filecolor=black,menucolor=black,urlcolor=brown}

 \def\su{\circleddash}
 \usepackage[all]{xy}

 \def\Li#1(#2){\textrm{Li}_{#1}\left(#2\right)}

\usepackage[compat=1.1.0]{tikz-feynman}
\usepackage{tikz,contour}
\usetikzlibrary{calc,decorations.markings,positioning}
\usepackage{pgfplots}
\pgfdeclarelayer{bg}    
\pgfsetlayers{bg,main}  

\newcommand{\DBoxNP}{	\begin{tikzpicture}[scale=0.05]
  \draw [black] (-2,1.5) to (0,1.5);
          		\draw [black] (-2,-1.5) to (0,-1.5);
		\draw [black] (-1,1.5) to (-1,-1.5);
		\draw [black] (0,1.5) to (2.5,-1.5);
		\draw [black] (0,-1.5) to (1.20,-0.06);
		\draw [black] (1.30,0.06) to (2.5,1.5);
                \draw [black] (0,1.5) to (2.5,1.5);
		\draw [black] (0,-1.5) to (2.5,-1.5);
		\draw [black] (2.5,-1.5) to (3.0,-1.5);
		\draw [black] (2.5,1.5) to (3.0,1.5);
              \end{tikzpicture}}

\newcommand{\IceCream}{	\begin{tikzpicture}[scale=0.05]
	\draw[color = black, fill=none] (2,0) arc(0:180:2);
		\draw [black] (-2,0) to (2,0);
		\draw [black] (-2,0) to (0,-2);
                	\draw [black] (2,0) to (0,-2);
	\end{tikzpicture}}

            \newcommand{\Kite}{	\begin{tikzpicture}[scale=0.05]
              		\draw [black] (-2,0) to (-0,1.5);
		\draw [black] (-2,0) to (-0,-1.5);
                	\draw [black] (2,0) to (0,1.5);
                        \draw [black] (2,0) to (0,-1.5);
                        \draw[black] (0,1.5) to (0,-1.5);\end{tikzpicture}}

\newcommand{\chickenDiag}{	\begin{tikzpicture}[scale=0.05]
		\filldraw [black] (-2,0) circle (2pt);
		\draw [black] (-2,0) to (-0,1.5);
		\draw [black] (-2,0) to (-0,-1.5);
		\draw [black] (0,1.5) to (2.5,-1.5);
		\draw [black] (0,-1.5) to (1.20,-0.06);
		\draw [black] (1.30,0.06) to (2.5,1.5);
		\draw [black] (0,1.5) to (2.5,1.5);
		\draw [black] (0,-1.5) to (2.5,-1.5);
\end{tikzpicture}}
\begin{document}
\preprint{IPHT-t23/094, LAPTH-003/24}
\title{\bf Algorithm for differential equations for Feynman integrals
  in general dimensions}
\author[]{Leonardo de la Cruz}
\author[]{and Pierre Vanhove}
\affiliation[]{Institut de Physique Th\'eorique, Universit\'e Paris-Saclay,
CEA, CNRS, F-91191 Gif-sur-Yvette Cedex, France}

 \abstract{
We present an algorithm for
determining the minimal order differential equations associated to a given
Feynman integral in dimensional or analytic regularisation.  The algorithm is an extension
of the Griffiths-Dwork pole reduction adapted to the case of twisted
differential forms.
In dimensional regularisation,  we demonstrate the applicability of this algorithm by explicitly providing the
inhomogeneous differential equations for the
multiloop two-point sunset integrals: up to 20 loops for
the equal-mass case, the generic mass case at two-  and 
three-loop orders. Additionally, we derive  the differential  operators
for various infrared-divergent two-loop graphs. In the analytic regularisation case, we apply our algorithm for deriving a system of partial  differential equations for 
regulated Witten diagrams, which   arise in the evaluation of  cosmological correlators of conformally coupled $\phi^4$
  theory in four-dimensional de Sitter space.

}
\maketitle

\section{Introduction}

Feynman integrals are  key ingredients in various areas of physics, and their accurate calculation, whether analytically or numerically, remains a significant hurdle in advancing our understanding of physical phenomena.
In particular, identifying the specific types of special functions
required to evaluate Feynman integrals has been an ongoing challenge
since the early days of quantum field theory~\cite{Golubeva,Pham} and
continues to be an active research field as recently reviewed, for instance
in~\cite{Panzer:2015ida,Duhr:2019wtr,Mizera:2019ose,Travaglini:2022uwo,Weinzierl:2022eaz,Badger:2023eqz}.

\medskip
The set of differential operators acting on a Feynman integral gives
important information about its analytic nature. 
Moreover, the differential equation is important for
evaluating physical observables by solving the system of
differential equations associated with Feynman integrals, either
analytically, in perturbation with respect to the kinematic
parameters or numerically. 
For instance, 
the differential  operator has real singularities at the position of
thresholds and pseudo-thresholds, and the order of the differential
operator is connected to the underlying algebraic geometry of the singular locus
of the integrand~\cite{Doran:2023yzu}. Intriguingly, there is  growing evidence suggesting that certain
Feynman integrals correspond to relative period integrals of singular
Calabi--Yau geometries, a connection explored in a number
of studies,
including~\cite{Brown:2009ta,Bloch:2014qca,Bloch:2016izu,Bourjaily:2018ycu,Bourjaily:2019hmc,Bourjaily:2018yfy,Klemm:2019dbm,Bonisch:2020qmm,Bonisch:2021yfw,Bourjaily:2022bwx,Forum:2022lpz,Duhr:2022pch,Frellesvig:2023PM,Pogel:2023zyd,Klemm:2024wtd}.
In addition, it has been remarked that correlation
functions~\cite{Heckelbacher:2022hbq}
and cosmological correlators~\cite{Heckelbacher:2022fbx,Chowdhury:2023arc}
of a conformally
coupled $\phi^4$ field in four dimensions can be expressed in terms of
flat space Feynman integrals. The regulation of ultraviolet
divergences in this case leads to integrals in analytical
regularisation.

In this work, we give an algorithmic procedure for deriving such
differential equations and the inhomogeneous part without having to go through the integral
reduction to master integrals and the construction of a reducible
system of differential equations satisfied by the set of master
integrals. Among the motivations for finding a shortcut  to derive differential equations without relaying on   master
integrals reduction is that the integration-by-parts reduction leads to large
system of master integrals that may obscure the algebraic
geometry underlying the analytic
structure of Feynman integrals.
Another motivation is the application to
cosmological correlators which give rise to analytic regularisation
for which the commonly used integration by part algorithms are not developed out-of-the-box. Finding a system of partial differential equations (PDEs) is also useful for generalised Feynman integrals in the context of 
Gel'fand-Kapranov-Zelevinski\u\i {} (GKZ) systems, which gives a D-module of differential operators 
acting on
the Feynman
integral~\cite{Vanhove:2018mto,delaCruz:2019skx,Klausen:2019hrg,Feng:2019bdx,Klemm:2019dbm,Ananthanarayan:2022ntm,Agostini:2022cgv,Matsubara-Heo:2023ylc,Munch:2022ouq}. However, the transition of this D-module to the PDEs of  Feynman integrals requires a restriction which is highly non-trivial and still an open problem \cite{delaCruz:2019skx,Klausen:2021yrt,Chestnov:2023kww,Dlapa:2023cvx}.

We work with the regularised parametric representation of a Feynman integral $
I_\Gamma^{\epsilon,\kappa}=\int_{x_i\geq0}  \omega_\Gamma^{\epsilon,\kappa} dx_1\cdots dx_n$ attached to a
graph $\Gamma$ (we refer to 
Section~\ref{sec:parametric} for details)
with
\begin{equation}\label{e:OmegaAnRegIntro}
  \omega^{\epsilon,\kappa}_\Gamma = { \textbf{U}_\Gamma ^{\nu_1+\cdots+\nu_n-(L+1)D}  \over \textbf{F}_\Gamma ^{\nu_1+\cdots+\nu_n-LD} }\, \prod_{i=1}^n x_i^{\nu_i-1} ,
\end{equation}
with  $D=2\delta-2\epsilon$ with $\delta$ a positive integer for
dimensional regularisation and $\nu_i=\upnu_i+\mu_i \kappa$ for
analytic regularisation.
In the case when $\epsilon=\kappa=0$ and $\delta$ a positive integer, the exponents
in~\eqref{e:OmegaAnRegIntro} are
integers and
we have a rational differential form. One may then use the  Griffiths-Dwork
pole
reduction~\cite{Griffiths_1969,Griffith1,Griffith2,Dwork_1962,Dwork_1964}
applied to the case of Feynman
  integrals~\cite{Golubeva,Muller-Stach:2011qkg,muller2014picard,Bloch:2016izu,Vanhove:2018mto,Lairez:2022zkj}
  for determining the minimal order differential operators associated with a given Feynman
integral. In integer space-time dimensions and without analytic
regulator, the integrand of the Feynman integral is a rational
differential form to which one can apply the generalised Griffith-Dwork algorithm~\cite{Lairez:2022zkj}.  When working in dimensional
regularisation, i.e. $\epsilon\neq0$, or analytic regularisation,
i.e. $\kappa\neq0$,  the integrand is a  twisted differential
form. One possible approach is a  direct application of the
Griffiths-Dwork pole
reduction~\cite{Muller-Stach:2011qkg,muller2014picard} or the creative
telescoping
algorithm~\cite{Chyzak,Chyzak2,bostan2013creative,Koutchan} but this
approach leads to large linear systems limiting its use for Feynman
graphs with many legs or many loops. 
Therefore, in  this work, we give an extension of the Griffiths-Dwork
reduction  algorithm which  make an essential use of the fact that the
twist
is build from the Symanzik graph polynomials $\textbf{U}_\Gamma$ and
$\textbf{F}_\Gamma$. This reduces the size of the linear system to be
solved for determining the   coefficients of
the differential operator. Because this
linear system is generically dense and of large size, we use the
finite field package \texttt{FiniteFlow}~\cite{Peraro:2019svx} to
derive analytic solutions.

 This way we
can analyse how the space-time dimension or the analytic regulator  affects the minimal order of the
differential operators.

This paper is organised as follows. 
In Section~\ref{sec:parametric}, we review the parametric
representation of Feynman integrals  setting our notation for the 
dimensionally and analytically regulated integrals in Section~\ref{sec:Twisted}.
In Section~\ref{sec:Red}, we
present the algorithmic procedure for deriving the differential
equations. In Section~\ref{sec:griff-dwork-reduct},  we generalise the
Griffiths-Dwork pole reduction to the case of the twisted differential
$\omega_\Gamma^{\epsilon,\kappa}$, and explain in
Section~\ref{sec:deriv-diff-equat} how to use this iteratively to
determine the differential equations. This generalises the algorithm for the rational differential form cases
used in~\cite{Lairez:2022zkj} to the case of twisted differential
forms appearing in Feynman integrals  in general dimensions. In Section~\ref{sec:oneloop}, we
illustrate the procedure by working on  the dimensionally regulated massless box.
We then consider the Witten cross diagram in (A)dS$_4$ in Section~\ref{sec:WittenCross}.
In
Section~\ref{sec:twoloop}, we derive the $\epsilon$-deformed
differential equation for the two-loop sunset integral for various
mass configurations, the massless double-box and the ice-cream cone
graph. We give a Gr\"obner basis of differential operators for the
analytically regulated
two-loop ice-cream graph, which arises in the cosmology correlator in dS$_4$. In Section~\ref{sec:higherloop}, we give the $\epsilon$-deformed
differential
equation for the equal-mass sunset up to twenty loop orders and for
the three-loop massive sunset.
In Section~\ref{sec:minim-order-diff},
we discuss the question of the minimal order differential operator,
and in Section~\ref{sec:epsilon-dependence} we analyse how the
$\epsilon$ parameter arises in the differential
equation. Section~\ref{sec:conclusion} contains a short
conclusion. The Appendix~\ref{sec:bessel} is dedicated to a short
discussion of the derivation of the differential equation using the
Bessel representation with the creative telescoping algorithm.

\section{Twisted differential from regulated Feynman integrals}
\label{sec:parametric}

We will apply our formalism in dimensional regularisation for Feynman integrals and analytic regularisation for Witten diagrams. Their treatment is slightly different, so we will consider them separately. 

\subsection{Review of the parametric representation}
We consider the parametric  representation of Feynman
integrals  in $D$ space-time dimensions associated with a graph $\Gamma$ with $n$   internal edges and $L$ loops. Its derivation can be found for instance in~\cite{nakanishi1971graph,Vanhove:2014wqa,Bogner:2010kv,Weinzierl:2022eaz}. 
The differential form associated to a Feynman integral is 
\begin{equation}\label{e:OmegaDef}
	\omega_\Gamma ={ \textbf{U}_\Gamma ^{\nu_1+\cdots+\nu_n-{(L+1)D\over2}}  \over
  \textbf{F}_\Gamma ^{\nu_1+\cdots+\nu_n-{LD\over2}} }\, \prod_{i=1}^n
x_i^{\nu_i-1}; \qquad 	\Omega_\Gamma :=	\omega_\Gamma \, \Omega^{(n)}_0,
\end{equation}
where we have introduced the  canonical differential form on
$\mathbb P^{n-1}$ 
\begin{equation}
	\Omega_0^{(n)}:=  \sum_{i=1}^n (-1)^{i-1} x_i dx_1\wedge \cdots \wedge \widehat{dx_i} \wedge\cdots \wedge dx_n ,
\end{equation}
and where  $\widehat{dx_i}$ means that the term is omitted in the wedge
product. We denote collectively the 
variables 
attached to all edges of
$\Gamma$ as $\underline x:=\{x_i |1\leq i\leq n\}$. The exponents $\nu_i$ are the powers of propagators (internal edges) of $\Gamma$. 
The polynomials $\textbf{U}_\Gamma$ and $\textbf{F}_\Gamma$
associated to  $\Gamma$ are defined as
follows~\cite{nakanishi1971graph,Bogner:2010kv,Weinzierl:2022eaz}.  The {\em first
	Symanzik polynomial} is defined by
\begin{equation}\label{e:Udef}
	{\bf U}_\Gamma (\underline x)= \sum_{\mathsf{T} \in \substack{ \text{Spanning} \\ \text{ trees of } \Gamma}} x_{\mathsf{T}}\,, 
\end{equation}
where the sum is over all the spanning trees $\mathsf{T}$ of $\Gamma$,  i.e.  all sub-graphs ${\sf T}$ of $\Gamma$ which contain all
vertices of $\Gamma$ so that the first Betti number (i.e. the number
of loops) $b_1(\mathsf{T}) =0$ and the number of connected component
is 
$b_0(\mathsf{T})=1$. The
monomial is the product of the variables not in the spanning tree
$x_{\mathsf{T}} = \prod_{e\notin {\mathsf{T}}} x_e$.  This is a
homogeneous polynomial of 
$\deg(\textbf{U}_\Gamma(\underline x))=L$.
The {\em second 
	Symanzik polynomial}  $\textbf{F}_\Gamma$ is defined by
\begin{equation}\label{e:VFdef}
	{\bf V}_{\Gamma} (\underline x)= \sum_{\substack{ \text{Spanning} \\ \text{ 2-forests
				of } \Gamma}}s_{\mathsf{F}} x_{\mathsf{F}}
	,\qquad {\bf F}_{\Gamma}(\underline x) = {\bf
		U}_\Gamma(\underline x)\left( \sum_{e \in e(\Gamma)} m_e^2 x_e\right) - {\bf V}_{\Gamma}\,,
\end{equation}
where $x_{\mathsf{F}} =
\prod_{e \notin \mathsf{F}} x_i$ to each spanning 2-forest. A 2-forest
is a disjoint union of two sub-trees $\mathsf{F}=\mathsf{T}_1\cup \mathsf{T}_2$. 
This is a homogeneous polynomial of  
$\deg(\textbf{F}_\Gamma(\underline x))=L+1$.
With the dot denoting the scalar product on $\mathbb R^{1,D-1}$,  we define the invariant of $\mathsf T$ as $s_\mathsf{F} = \sum_{(v_1,v_2) \in \mathsf{F}=\mathsf{T}_1\cup \mathsf{T}_2} p_{v_1}\cdot p_{v_2}$ .
The second Symanzik polynomial carries all dependence on the physical parameters, that is 
the
internal masses and the external kinematics, which we  write as 
\begin{equation}
	\vec m:=\{m_1,\dots,m_n\}, \quad  \vec s:=\{s_\textsf{F} \mid \textsf{F} ~ \textrm{spanning
		2-forests of}~\Gamma\}\, ,	
\end{equation}
respectively.

The differential form~(\ref{e:OmegaDef}) is defined in the middle cohomology $H^{n-1}(\mathbb P^{n-1}\backslash
\{\textbf{U}_\Gamma \textbf{F}_\Gamma=0\})$~\cite{bek,Brown:2009ta}.
The Feynman integral associated with the graph $\Gamma$ is 
given by the integral $  I_\Gamma = \int_{\Delta_n}
\, \Omega_\Gamma$ of the  differential form over
the positive orthant
\begin{equation}\label{e:Deltan}
	\Delta_n:=\{(x_1,\dots,x_n)\in \mathbb P^{n-1}, x_i \geq0 ~\textrm{for}~ 1\leq
	i\leq n\} \, .
\end{equation}

\subsection{Twisted differential forms}\label{sec:Twisted}

The Feynman integral $I_\Gamma$ may diverge  for integer values of $D$ and 
the exponents $\nu_i$, but there is an open subset of
$(D,\nu_1,\dots,\nu_n)\in\mathbb C^{n+1}$ where  the integral
converges. The (unique) value of the Feynman integral is  defined by analytic continuation~\cite{Speer}.
We work in dimension $D=2\delta-2\epsilon$ with $\delta\in\mathbb N^*$ and
$\epsilon\in\mathbb R$ and  consider as well the situation where the powers of the propagators are shifted from integer values, that is  $\nu_i=\upnu_i
+ p_i \kappa$ with $(\upnu_1,\dots,\upnu_n,p_1,\dots,p_n)\in\mathbb Z^{2n}$. The 
differential form~(\ref{e:OmegaDef}) thus becomes a \emph{twisted
	differential form}  $\Omega^{\epsilon,\kappa}_\Gamma
=\omega^{\epsilon,\kappa}_\Gamma \Omega_0^{(n)}$ with
\begin{equation}\label{e:OmegaTwistAnaReg}
	\omega^{\epsilon,\kappa}_\Gamma := { \textbf{U}_\Gamma
		^{\upnu_1+\cdots+\upnu_n-(L+1)\delta}  \over \textbf{F}_\Gamma
		^{\upnu_1+\cdots+\upnu_n-L\delta}
	}\,\left(\textbf{U}_\Gamma^{L+1}\over \textbf{F}_\Gamma^L\right)^\epsilon
	\left(\prod_{i=1}^n \left(x_i \textbf{U}_\Gamma(\underline x)\over
	\textbf{F}_\Gamma(\underline x)\right)^{p_i}\right)^{\kappa} \, \prod_{i=1}^n x_i^{\upnu_i-1} \,.
\end{equation}
The twists are the $\epsilon$ or $\kappa$-th powers of  homogeneous
degree zero rational functions on $\mathbb
P^{n-1}$.

Notice that we do not assume that $\epsilon$ or $\kappa$ are small numbers.
When $\kappa=0$ we have the parametric representation of a Feynman
integral in dimensional regularisation, and we will use the short-hand
notation 
\begin{equation}\label{e:OmegaTwistDimReg}
\omega^\epsilon_\Gamma:=\omega^{\epsilon,0}_\Gamma = { \textbf{U}_\Gamma ^{\nu_1+\cdots+\nu_n-(L+1)\delta}  \over \textbf{F}_\Gamma ^{\nu_1+\cdots+\nu_n-L\delta} }\,
	\left(\textbf{U}_\Gamma^{L+1}\over \textbf{F}_\Gamma^{L}\right)^{\epsilon} \, \prod_{i=1}^n x_i^{\nu_i-1} \, .
\end{equation}

\medskip

The differential forms~(\ref{e:OmegaTwistDimReg})
and~(\ref{e:OmegaTwistAnaReg}) are twisted differential of the kind
studied in~\cite{Aomoto1,Aomoto,Aomoto_1982,AomotoBook}.
Their relevance to Feynman integrals was already recognised in these works and has
been applied in
e.g.~\cite{Mizera:2017rqa,Frellesvig:2019uqt,Caron-Huot:2021xqj,Caron-Huot:2021iev,Cacciatori:2021nli,Fontana:2023amt,Munch:2023ifm,Brunello:2023rpq,De:2023xue,Teschke:2024bct}  for
expanding Feynman integrals on the basis of master integrals. In contrast to these works, 
we will use
the fact that the twist is given by
the power of the homogeneous degree 0 rational form
which will be essential
in the construction presented in this work.

\section{Annihilators of Feynman integrals}
\label{sec:Red}
Feynman integrals are holonomic functions of their physical parameters~\cite{Kashiwara:1977nf, Bitoun:2017nre, Smirnov:2010hn,Lee:2013hzt}. This means that Feynman integrals satisfy systems of  (inhomogeneous) partial
differential equations of finite order with respect to their physical
parameters $\vec m$ and $\vec s$.

Let us consider $r$ parameters from the set of internal masses and independent kinematics, $\underline z:=\{z_1,\dots,z_r\} \in \vec m \cup \vec s $. 
We seek 
differential operators 
annihilating the differential form $\Omega_\Gamma^{\epsilon,\kappa}$ 
\begin{equation}\label{e:PFOmegaGeneric}
	\left(  \sum_{a_1=0}^{o_1} \sum_{ a_r=0}^{o_r} c_{a_1,\dots,a_r}(\vec m,\vec s,\epsilon,\kappa)\left(\partial\over \partial z_1\right)^{a_1}\cdots\left(\partial
	\over\partial z_r\right)^{a_r}  \right)\Omega_\Gamma^{\epsilon,\kappa}=d\beta^{\epsilon,\kappa}_\Gamma,
\end{equation}
where $c_{a_1,\dots,a_r}(\vec m,\vec s,\epsilon,\kappa)$ are rational functions of the physical
parameters, but they are independent of the edge variables $x_1,\dots,x_n$. The inhomogeneous term is a total derivative in $x_i$'s where the only allowed poles are those already present in $\Omega_\Gamma^{\epsilon,\kappa}$~\cite{Lairez:2022zkj}.
Because the domain of integration of the Feynman integral does not
depend on the physical parameters, we then deduce
\begin{equation}\label{e:PDEintFeynman}
	\left( \sum_{a_1=0}^{o_1} \sum_{ a_r=0}^{o_r} c_{a_1,\dots,a_r}(\vec m,\vec s,\epsilon,\kappa)\left(\partial\over \partial z_1\right)^{a_1}\cdots\left(\partial
	\over\partial z_r\right)^{a_r}  \right)I^{\epsilon,\kappa}_\Gamma=\mathscr{S}^{\epsilon,\kappa}_\Gamma \, ,
\end{equation}
where $\mathscr{S}^{\epsilon,\kappa}_\Gamma$ is an inhomogeneous term obtained by
integrating $d\beta^{\epsilon,\kappa}_\Gamma$ over the boundary of
orthant~\eqref{e:Deltan}. This is a non-trivial task because one needs
to blow-up the intersections between the graph hypersurface and the
domain of integration, so the integral is well-defined~\cite{bek,Brown:2009ta,Bloch:2016izu,muller2014picard}. For
instance, 
Section~3.2 of~\cite{Bloch:2016izu} gives  a detailed derivation of the inhomogeneous term for
the two-loop sunset integral along these lines.
If the integration is done over a cycle $\mathcal C$, like the one
defined  by the maximal cut $\mathcal C_{\rm
  max}:=\{|x_1|=\cdots=|x_n|=1\}$,  the resulting integral is
annihilated by the  action of the differential operator~\cite{Vanhove:2018mto}
\begin{equation}
	\left( \sum_{a_1=0}^{o_1} \sum_{ a_r=0}^{o_r} c_{a_1,\dots,a_r}(\vec m,\vec s,\epsilon,\kappa)\left(\partial\over \partial z_1\right)^{a_1}\cdots\left(\partial
	\over\partial z_r\right)^{a_r}  \right)\int_{\mathcal C }\Omega^{\epsilon,\kappa}_\Gamma=0\,.
\end{equation}

The ideal  generated by these
differential operators is a differential module (or D-module).  Thus, the
differential equations we are seeking can be obtained by deriving  annihilators  of
$\Omega_\Gamma^{\epsilon,\kappa}$, i.e., partial differential operators that
annihilate  the integrand by acting on the physical parameter and the
edge variables.

\subsection{Griffiths-Dwork reduction for twisted differential forms}\label{sec:griff-dwork-reduct}
The    differentiation of $\Omega_\Gamma^{\epsilon,\kappa}$ 
leads  to  expressions of the type 
\begin{multline}\label{e:PDE}
\sum_{\mathsf{a}=a_1+\cdots+a_r\atop
  a_i\geq0}c_{a_1,\dots,a_r}(\vec m,\vec s,\epsilon,\kappa)\left(\partial\over \partial z_1\right)^{a_1}\cdots\left(\partial
  \over\partial z_r\right)^{a_r} \Omega_\Gamma^{\epsilon,\kappa}=\cr \sum_{\mathsf{a}=a_1+\cdots+a_r\atop a_i\geq0}
{ 
  c_{a_1,\dots,a_r}(\vec m,\vec s,\epsilon,\kappa)  P^{(a_1,\dots,a_r)}(\underline x)\over \textbf{F}_\Gamma^{\mathsf {a}}}\, \Omega_\Gamma^{\epsilon,\kappa},
\end{multline}
where 
$  P^{(a_1,\dots,a_r)}(\underline x)$ is a
  homogeneous polynomial of degree $(L+1)(a_1+\cdots+a_r)$ in
  the edge variables $\underline x$.
  The sum is over the differential operators of order
  $a_1\geq0 ,\dots, a_r\geq0$ and fixed total order $\mathsf{a}:=a_1+\cdots +a_r$. 
The
  pole order in the second Symanzik polynomial $\textbf{F}_\Gamma$ has increased by
  $\mathsf a$. 
  To derive~Eq.~\eqref{e:PFOmegaGeneric} one needs to find the
  coefficient $ c_{a_1,\dots,a_r}(\vec m,\vec s,\epsilon,\kappa)$.  From now on we consider the case where
  $\nu_1=\cdots=\nu_r=1$ so that $\nu=n$. The case with $\nu_i\neq1$
  is an immediate generalisation.

  \subsubsection{The pole reduction for dimensional
    regularisation}\label{sec:PoleRed}
We adapt the Griffiths-Dwork pole reduction to the case
of the twisted differential form~(\ref{e:OmegaTwistDimReg}) in
dimensional regularisation (i.e. $\kappa=0$ and $\epsilon\neq0$).
The starting point of the algorithm is the reduction of polynomial
$P^{(a_1,\dots,a_r)}(\underline x)$  in the numerator
of~\eqref{e:PDE}
\begin{equation}\label{e:RedF}
	P^{(a_1,\dots,a_r)}(\underline x) = \vec C_{\mathsf{a}}(\underline x)\cdot
	\vec\nabla   \textbf{F}_\Gamma \, ,
      \end{equation}
      where we have introduced the gradient $	\vec\nabla   \textbf{F}_\Gamma :=\left(\partial_{x_1}
      \textbf{F}_\Gamma(\underline x),\dots, \partial_{x_n}
      \textbf{F}_\Gamma(\underline x)\right)$.
The components of the size $n$ vector $ \vec C_{\mathsf{a}}(\underline x)$ are homogeneous polynomials of degree
$\mathsf{a}(L+1)-L$ in  $\underline x$. 
   We generalise the construction by
   Griffiths~\cite{Griffith1,Griffith2} to include the twist factor
   for $\mathsf{a}>1$ 
   \begin{equation}\label{e:betadef}
  \beta^{(a_1,\dots,a_r)}=  \sum_{1\leq i<j\leq n} {x_i
    G_{\mathsf{a}}^j  (\underline x)-x_j
   G_{\mathsf{a}}^i(\underline x)\over
  \textbf{F}_\Gamma^{\mathsf{a}-1}}\, 
 dx_1\wedge \cdots \wedge \widehat{dx_i}\wedge \cdots \wedge\widehat{dx_j}\wedge
  \cdots \wedge dx_n \, .
\end{equation}
To take into account the general dimensional case, we have introduced 
the vectors of twisted forms
\begin{equation}
  \label{e:Gdef}
\vec  G_{\mathsf{a}}(\underline x):=   \vec C_{\mathsf{a}} \,\omega_\Gamma^\epsilon \, ,
\end{equation}
whose components are of homogeneous degree $(\mathsf{a}-1)(L+1)+1-n$. 
Following the same steps as in~\cite{Griffiths_1969}, we have
\begin{multline}
  d\beta_\Gamma^{(a_1,\dots,a_r)}=-  (\mathsf{a}-1) \sum_{1\leq i<j\leq n} {x_i
    G_{\mathsf{a}}^j  (\underline x)-x_j
   G_{\mathsf{a}}^i(\underline x)\over
   \textbf{F}_\Gamma^{\mathsf{a}}}\, d \textbf{F}_\Gamma\wedge
 dx_1\wedge \cdots \wedge \widehat{dx_i}\wedge \cdots \wedge\widehat{dx_j}\wedge
 \cdots \wedge dx_n \cr
+  \sum_{1\leq i<j\leq n} {d(x_i
    G_{\mathsf{a}}^j  (\underline x)-x_j
   G_{\mathsf{a}}^i(\underline x))\over
   \textbf{F}_\Gamma^{\mathsf{a}-1}}\wedge
 dx_1\wedge \cdots \wedge \widehat{dx_i}\wedge \cdots \wedge\widehat{dx_j}\wedge
  \cdots \wedge dx_n  \, .
\end{multline}
From the degree of homogeneity of $\textbf{F}_\Gamma$ and the
components of $\vec G_{\mathsf{a}}(\underline x)$ 
\begin{align}
  \sum_{i=1}^n x_i {\partial\textbf{F}_\Gamma (\underline x)\over \partial x_i}&=
                                                                  (L+1) \textbf{F}_\Gamma (\underline x),\cr
   \sum_{i=1}^n x_i {\partial\vec G_{\mathsf{a}}(\underline x)\over \partial x_i}&=
                                                                  ((\mathsf{a}-1)(L+1)+1-n) \vec G_{\mathsf{a}}(\underline x) \,,
\end{align}
we find that 
   \begin{equation}
     d\beta_\Gamma^{(a_1,\dots,a_r)}= (\mathsf{a}-1) {  \vec  G_{\mathsf{a}}
     (\underline x)
\cdot    \vec \nabla\textbf{F}_\Gamma\over
     \textbf{F}_\Gamma^{\mathsf{a}}}\Omega_0^{(n)}-{
\vec\nabla \cdot \vec G_{\mathsf{a}}
     (\underline x)
    \over
   \textbf{F}_\Gamma^{\mathsf{a}-1}}
  \Omega_0^{(n)} \,.
\end{equation}
Using the definition of $\vec  G_{\mathsf{a}}$ in~\eqref{e:Gdef} we
have reduced the pole order  of $ \textbf{F}_\Gamma$ in~\eqref{e:PDE} 
\begin{equation}\label{e:RedG1}
 (\mathsf{a}-1)\left(\partial\over \partial z_1\right)^{a_1}\cdots\left(\partial
  \over\partial z_r\right)^{a_r} \Omega_\Gamma^\epsilon={
\vec\nabla \cdot \vec G_{\mathsf{a}}
     (\underline x)
    \over
  \textbf{F}_\Gamma^{\mathsf{a}-1}}
  \Omega_0^{(n)}+d\beta_\Gamma^{(a_1,\dots,a_r)}.
\end{equation}
We now expand the first term in the right-hand-side
\begin{equation}
  \vec\nabla \cdot \vec G_{\mathsf{a}}(\underline x)=  \vec\nabla \cdot \vec C_{\mathsf{a}}(\underline x) {\textbf{U}_\Gamma^{\lambda_U}\over
  \textbf{F}_\Gamma^{\lambda_F}} 
+
  \vec C_{\mathsf{a}}(\underline x) \cdot\vec\nabla  \left(   {\textbf{U}_\Gamma^{\lambda_U}\over
  \textbf{F}_\Gamma^{\lambda_F}} \right),
\end{equation}
where we have defined
\begin{equation}\label{e:powerUFDef}
  \lambda_U=n-(L+1)(\delta-\epsilon), \qquad 
  \lambda_F=n-L(\delta-\epsilon).
  \end{equation}
The second term in this equation can be evaluated using 
\begin{align}
  \vec C_{\mathsf{a}}(\underline x)\cdot \vec\nabla \left(\textbf{U}_\Gamma^{\lambda_U}\over
\textbf{F}_\Gamma^{\lambda_F}\right)&=\left(\lambda_U \vec
C_{\mathsf{a}}\cdot \vec\nabla \log\textbf{U}_\Gamma -{\lambda_F} \vec
C_{\mathsf{a}}\cdot \vec\nabla \log \textbf{F}_\Gamma\right) {\textbf{U}_\Gamma^{\lambda_U}\over
\textbf{F}_\Gamma^{\lambda_F}} \cr
&=\left[-{\lambda_F}{
    P^{(a_1,\dots,a_r)}(\underline x)\over
  \textbf{F}_\Gamma}   
+
 \lambda_U \vec
C_{\mathsf{a}}\cdot \vec\nabla \log\textbf{U}_\Gamma \right] {\textbf{U}_\Gamma^{\lambda_U}\over
	\textbf{F}_\Gamma^{\lambda_F}}\,,
\end{align}
where we have used Eq.~\eqref{e:RedF} in the second equality. Therefore, 
\begin{equation}
	\label{e:Pred}
\left(\partial\over \partial z_1\right)^{a_1}\cdots\left(\partial
  \over\partial z_r\right)^{a_r} \Omega_\Gamma^\epsilon={
\vec\nabla \cdot \vec C_{\mathsf{a}}
     (\underline x)
+ \lambda_U \vec
    C_{\mathsf{a}}\cdot\vec\nabla\log\textbf{U}_\Gamma\over \left(\mathsf{a}-1+\lambda_F\right)\,\textbf{F}_\Gamma^{\mathsf{a}-1}}
  \Omega_\Gamma^\epsilon
+{  1\over \mathsf{a}-1+\lambda_F} d\beta_\Gamma^{(a_1,\dots,a_r)}.
\end{equation}
This expression involves the term $\vec
    C_{\mathsf{a}}\cdot\vec\nabla\log\textbf{U}_\Gamma$ which has a
    pole in $\textbf{U}_\Gamma$. We then perform a second reduction by demanding that
    \begin{equation}
      \label{e:RedU}
      \vec
    C_{\mathsf{a}}(\underline x)\cdot\vec\nabla \textbf{U}_\Gamma =
    c_{\mathsf{a}}(\underline x) \, \textbf{U}_\Gamma \, ,
    \end{equation}
where $ c_{\mathsf{a}}(\underline x)$ is a homogeneous polynomial of
degree $(\mathsf{a}-1)(L+1)$. This is equivalent to the computation of
syzygies of $\text{Jac}( \textbf
U_\Gamma):=\langle \vec\nabla \textbf{U}_\Gamma(\underline x)\rangle$. Indeed,  using the homogeneity of $\textbf{U}_\Gamma$ we can rewrite the previous equation as 
\begin{equation}
 \left( L   \vec
C_{\mathsf{a}}(\underline x) -
c_{\mathsf{a}}(\underline x) \vec{x} \right) \, \cdot\vec\nabla
\textbf{U}_\Gamma =0\, ,
\end{equation}
which are examples of syzygies of the Jacobian of $\textbf U_\Gamma$. Using this reduction in Eq.~\eqref{e:Pred} leads to
\begin{equation}\label{e:RedFinal}
\left(\partial\over \partial z_1\right)^{a_1}\cdots\left(\partial
  \over\partial z_r\right)^{a_r} \Omega_\Gamma^\epsilon={
M^{(a_1,\dots,a_r)}
     (\underline x)
\over \textbf{F}_\Gamma^{\mathsf{a}-1}}\Omega_\Gamma^\epsilon
+{  \mathsf{a}-1\over \mathsf{a}+n-L(\delta-\epsilon)} d\beta_\Gamma^{(a_1,\dots,a_r)}
\end{equation}
with the numerator given by the polynomial of homogeneous degree $(\mathsf{a}-1)(L+1)$
\begin{equation}
  \label{e:Mdimregfinal}
  M^{(a_1,\dots,a_r)}(\underline x):={\vec\nabla \cdot \vec C_{\mathsf{a}}
     (\underline x)
+ \lambda_U\, 
    c_{\mathsf{a}}(\underline x)\over \mathsf{a}-1+\lambda_F },
\end{equation}
with $\lambda_U$ and $\lambda_F$ the powers of the $\textbf{U}_\Gamma$
and the $\textbf{F}_\Gamma$ polynomials respectively given in~\eqref{e:powerUFDef}.

To perform the pole reduction, 
 we have to solve the linear system
\begin{equation}\label{e:sysCFU}
   \left\{\begin{array}{@{}l@{}}
\vec C_{\mathsf{a}} (\underline x) \cdot \vec \nabla \textbf{F}_\Gamma
            =    P^{(a_1,\dots,a_r)}(\underline x)\\
\vec C_{\mathsf{a}} (\underline x)\cdot\vec \nabla\textbf{U}_\Gamma= c_{{\mathsf{a}}}(\underline x) \textbf{U}_\Gamma 
  \end{array}\right.\,,
\end{equation}
for determining the coefficients of $\vec C_{\mathsf{a}}(\underline
x)$ and $c_{\mathsf{a}}(\underline x)$.
The system~\eqref{e:sysCFU} has a solution when its rank is positive.
We have a linear system of the $n$ components of $\vec C_{\mathsf{a}}(\underline
x) $ which are homogeneous polynomial of degree $\deg(C)=\deg( P^{(a_1,\dots,a_r)})-L $ in
$\underline x$ and  $ c_{{\mathsf{a}}}(\underline x) $ which is a polynomial of
homogeneous degree $\deg(C)-1$.   Since the number of coefficients of
a homogeneous polynomial of degree $d$ in $n$ variables is
$\binom{d+n-1}{d}$,  the system has
\begin{equation}\label{e:nvars}
n \binom{\deg( P^{(a_1,\dots,a_r)})-L+n-1}{\deg( P^{(a_1,\dots,a_r)})-L}+\binom{\deg( P^{(a_1,\dots,a_r)})-L+n-2}{\deg( P^{(a_1,\dots,a_r)})-L-1}
\end{equation} unknown variables for
\begin{equation}\label{e:nequ}
\binom{\deg( P^{(a_1,\dots,a_r)})+n-1}{\deg( P^{(a_1,\dots,a_r)})}+\binom{\deg( P^{(a_1,\dots,a_r)})+n-2}{\deg( P^{(a_1,\dots,a_r)})-1}
\end{equation}
equations.
Since the $\deg(P^{(a_1,\dots,a_r)})=\mathsf{a}(L+1)$, the rank of the system~\eqref{e:sysCFU} is
  \begin{align}
    \label{e:Rank}
    \textrm{rank}&=\eqref{e:nvars}-\eqref{e:nequ}\\
\nonumber    &=n \binom{(L+1)(\mathsf{a}-1)+n
               }{(L+1) (\mathsf{a}-1)+1}+\binom{(L+1)(\mathsf{a}-1)+n-1}{(L+1) (\mathsf{a}-1)}\cr
             \nonumber    &
                              -\binom{(L+1) \mathsf{a}+n-1}{(L+1) \mathsf{a}}-\binom{(L+1) \mathsf{a}+n-2}{(L+1) \mathsf{a}-1}
  \end{align}
For fixed values of loops  $L$
and number of edges $n$ there is always a value of the
number of derivatives $\mathsf{a}$ such that the system has
positive rank.

A few comments are in order. In practice for Feynman integrals,   the polynomial $ P^{(a_1,\dots,a_r)}(\underline x)$ is not a generic
homogeneous polynomial, so the number of equations is smaller or equal than~\eqref{e:nequ}.
We remark that this way of solving the linear system includes implicitly the
freedom given by the syzygies of $\textrm{Jac}(\textbf{F}_\Gamma)
 :=\langle \vec\nabla \textbf{F}_\Gamma(\underline x)\rangle$ and
$\textrm{Jac}(\textbf{U}_\Gamma)$ since they belong to
the kernel of equation~\eqref{e:RedF} and~\eqref{e:RedU}
respectively.\footnote{It was noticed in~\cite{Lairez:2022zkj},  that in
  the rational case, only the first order syzygies are needed to take
  into account the non-isolated singularities of Feynman integrals.}
One important property of that reduction is that the differential form $\beta_\Gamma^{(a_1,\dots,a_n)}$ is that it
does not have poles that are not poles of $\textbf{F}_\Gamma$ which is
guaranteed by  construction. We refer
to Section~3 of~\cite{Lairez:2022zkj} for a discussion of the pole constraints.

  The system of linear equation~\eqref{e:sysCFU} is dense since, in general, all coefficients in 
	$\vec C_{\mathsf{a}}(\underline
	x)$ and $c_{\mathsf{a}}(\underline x)$ are non-vanishing.   Moreover we are interested in analytic solutions of these systems. We thus benefit from the dense solver implemented in the Mathematica package \texttt{
		FiniteFlow}, described in detail in Sec.4
              of~\cite{Peraro:2019svx}.   Specifically, in Mathematica
              have used the command \texttt{FFDenseSolve}.

%
%
  \subsubsection{The pole reduction for analytic
    regularisation}\label{sec:PoleRedAn}
We give an adaption of the Griffiths-Dwork pole reduction to the case
of the twisted differential form~(\ref{e:OmegaTwistAnaReg})  from
analytical regularisation. Since most of the steps are similar to the
one presented in the previous section we only give the main equations.

As before we reduce the  polynomial
$P^{(a_1,\dots,a_r)}(\underline x)$ in the Jacobian of
$\textbf{F}_\Gamma$ using Eq.~(\ref{e:RedF}) and introduce the
differential forms~\eqref{e:betadef}
$\beta_\Gamma^{(a_1,\dots,a_r)}$  and the vector of differential forms $\vec
G_{\mathsf{a}}(\underline x)$ as in Eq.~\eqref{e:Gdef}, leading to the
pole reduction Eq.~\eqref{e:RedG1}.
Because the twist is different the expansion of the 
right-hand-side  (recall that $D=2\delta-2\epsilon$)
\begin{equation}
  \vec\nabla \cdot \vec G_{\mathsf{a}}(\underline x)=  \vec\nabla \cdot \vec C_{\mathsf{a}}(\underline x) { \textbf{U}_\Gamma ^{\lambda_U}  \over \textbf{F}_\Gamma ^{\lambda_F} }\,
  Q(\underline x)^\kappa
+
  \vec C_{\mathsf{a}}(\underline x) \cdot\vec\nabla  \left({ \textbf{U}_\Gamma ^{\lambda_U}  \over \textbf{F}_\Gamma ^{\lambda_F} }\,
  Q(\underline x)^\kappa \right) ,
\end{equation}
where we have set (we have assumed that $\upnu_1=\cdots=\upnu_n=1$ the
generic case of integer values is an easy generalisation)
\begin{equation}
  Q(\underline x):=\prod_{i=1}^n x_i^{p_i}   
\end{equation}
and defined the powers of the various polynomials by
\begin{equation}
  \label{e:powerUFQDef}
  \lambda_U= n-(L+1)(\delta-\epsilon)+\kappa\sum_{i=1}^n p_i, \qquad
  \lambda_F= n-L(\delta-\epsilon)+\kappa\sum_{i=1}^n p_i, \qquad
  \lambda_Q  = \kappa\,.
\end{equation}
This is  evaluated using 
\begin{equation}
  \vec C_{\mathsf{a}}(\underline x)\cdot \vec\nabla
  \left(\textbf{U}_\Gamma^{\lambda_U}Q(\underline x)^\kappa\over
\textbf{F}_\Gamma^{\lambda_F}\right)=\Big(\lambda_U \vec
C_{\mathsf{a}}\cdot \vec\nabla \log\textbf{U}_\Gamma -\lambda_F \vec
C_{\mathsf{a}}\cdot \vec\nabla \log \textbf{F}_\Gamma+\lambda_Q \vec
C_{\mathsf{a}}\cdot\vec\nabla \log Q(\underline                             x)\Big)
                            {\textbf{U}_\Gamma^{\lambda_U}Q(\underline x)^\kappa\over
\textbf{F}_\Gamma^{\lambda_F}} ,
\end{equation}
so that
\begin{multline}
	\label{e:PredAnReg}
\left(\partial\over \partial z_1\right)^{a_1}\cdots\left(\partial
  \over\partial z_r\right)^{a_r} \Omega_\Gamma^{\epsilon,\kappa}={
\vec\nabla \cdot \vec C_{\mathsf{a}}
     (\underline x)
+  \lambda_U \vec
    C_{\mathsf{a}}\cdot\vec\nabla\log\textbf{U}_\Gamma+\lambda_Q \vec
    C_{\mathsf{a}}\cdot\vec\nabla\log Q(\underline x)\over
    \left(\mathsf{a}-1+\lambda_F\right)\,\textbf{F}_\Gamma^{\mathsf{a}-1}}\,
  \Omega_\Gamma^{\epsilon,\kappa}\cr
+{  1 \over \mathsf{a}-1+\lambda_F} d\beta_\Gamma^{(a_1,\dots,a_r)}.
\end{multline}
This time we need to reduce the pole in $\textbf{U}_\Gamma$ from  the term $\vec
    C_{\mathsf{a}}\cdot\vec\nabla\log\textbf{U}_\Gamma$ and the new
    pole in $1/x_i$ arising from the propagators.
    As before we  impose the following conditions
\begin{equation}\label{e:sysCFUAnReg}
   \left\{\begin{array}{@{}l@{}}
\vec C_{\mathsf{a}} (\underline x) \cdot \vec \nabla \textbf{F}_\Gamma
            =    P^{(a_1,\dots,a_r)}(\underline x)\\
\vec C_{\mathsf{a}} (\underline x)\cdot\vec \nabla\textbf{U}_\Gamma=
            c_{{\mathsf{a}}}(\underline x) \textbf{U}_\Gamma \\
            \vec C_{\mathsf{a}} (\underline x)\cdot\vec \nabla
            Q(\underline x)=
            q_{{\mathsf{a}}}(\underline x) Q(\underline x)
  \end{array}\right.\,,
\end{equation}
leading to the pole  reduction in Eq.~\eqref{e:PDE} 
\begin{equation}\label{e:RedFinalAnReg}
\left(\partial\over \partial z_1\right)^{a_1}\cdots\left(\partial
  \over\partial z_r\right)^{a_r} \Omega_\Gamma^\epsilon={
M^{(a_1,\dots,a_r)}
     (\underline x)
\over \textbf{F}_\Gamma^{\mathsf{a}-1}}\Omega_\Gamma^\epsilon
+{  1\over \mathsf{a}-1+\lambda_F} d\beta_\Gamma^{(a_1,\dots,a_r)}. 
\end{equation}
The numerator is now given by the polynomial of homogeneous degree $(\mathsf{a}-1)(L+1)$
\begin{equation}
  \label{e:MfinalAnReg}
  M^{(a_1,\dots,a_r)}(\underline x):={
\vec\nabla \cdot \vec C_{\mathsf{a}}
     (\underline x)
+  \lambda_U
    c_{\mathsf{a}}(\underline x)+\lambda_Q q_{\mathsf{a}}(\underline x)\over
    \mathsf{a}-1+\lambda_F}.
\end{equation}
%

\subsubsection{Determination of the differential equations}
\label{sec:deriv-diff-equat}

We turn to the derivation of the differential
equation~\eqref{e:PFOmegaGeneric}  by iterating the generalised
Griffiths-Dwork reduction given in the previous sections. We present an algorithm for a derivation of
a linear ordinary differential equation with respect to a single
variable differentiation $z$ (either an internal mass, or a kinematic
variable or a scaling parameter as used
in~\cite{Lairez:2022zkj,Doran:2023yzu}) so that $r=1$ and
$\mathsf{a}=a_1$. The generalisation to the many variable  case is immediate.

We are seeking the differential operator 
\begin{equation}\label{e:PFgeneric}
  \mathscr{L}_\Gamma^{\epsilon,\kappa}=\sum_{\mathsf{a}=0}^{N(\Gamma,\epsilon,\kappa)}
  c_{\mathsf{a}}(\vec m,\vec s,\epsilon,\kappa) \left(d\over dt\right)^{\mathsf{a}}
\end{equation}
with $c_{\mathsf{a}}(\vec m,\vec s,\epsilon,\kappa)$ polynomials in the internal masses
$\vec m$  and the (independent) kinematic variables $\vec s$  and the
regulators $\epsilon$  and $\kappa$, such that
\begin{equation}
     \mathscr{L}_\Gamma^{\epsilon,\kappa} \Omega_\Gamma^{\epsilon,\kappa}= d\beta^{\epsilon,\kappa}_\Gamma.
   \end{equation}
Holonomicity of Feynman integrals gives an upper bound on the order of the differential operator, which is  determined by
 the number of master
integrals. For a graph $\Gamma$ the minimal order of the differential operator depends in general of the regulators.
Let $N(\Gamma,\epsilon,\kappa)$  be the starting order of the
reduction.  We then apply the results of Sections~\ref{sec:PoleRed}
or~\ref{sec:PoleRedAn} so that
\begin{equation}\label{e:step1}
\left(d\over dt\right)^{N(\Gamma,\epsilon,\kappa)} \Omega_\Gamma^\epsilon= {M^{
      N(\Gamma,\epsilon,\kappa)}(\underline x)\over
    \textbf{F}_\Gamma^{N(\Gamma,\epsilon,\kappa)-1}
  } \,\Omega_\Gamma^{\epsilon,\kappa}+ d\beta_\Gamma^{N(\Gamma,\epsilon,\kappa)} . 
\end{equation}
In the next step we add the lowest-order derivative

\begin{multline}\label{e:step2}
\left[\left(d\over dt\right)^{N(\Gamma,\epsilon,\kappa)}
  +q_{N(\Gamma,\epsilon, \kappa)-1}(t,\epsilon,\kappa) \left(d\over dt\right)^{N(\Gamma,\epsilon,\kappa)-1}\right] \Omega_\Gamma^{\epsilon,\kappa}\cr= {M^{
      N(\Gamma,\epsilon,\kappa)}(\underline x)+q_{N(\Gamma,\epsilon,\kappa)-1}(t,\epsilon,\kappa)
    P^{N(\Gamma,\epsilon,\kappa)-1}(\underline x) \over
    \textbf{F}_\Gamma^{N(\Gamma,\epsilon,\kappa)-1}
  }\,\Omega_\Gamma^{\epsilon,\kappa}+ d\beta_\Gamma^{N(\Gamma,\epsilon,\kappa)}\, ,  
\end{multline}
where the rational coefficient
$q_{N(\Gamma,\epsilon,\kappa)-1}(t,\epsilon)=c_{N(\Gamma,\epsilon,\kappa)-1}(t,\epsilon)/c_{N(\Gamma,\epsilon,\kappa)}(t,\epsilon)$
is an unknown rational function of $t$ and  the regulators $\epsilon$
and $\kappa$. 
The polynomial $ P^{N(\Gamma,\epsilon,\kappa)-1}(\underline x) $
is the numerator factor obtained by taking the $N(\Gamma,\epsilon,\kappa)-1$
derivative of the differential form.

We then apply the reduction of Section~\ref{sec:PoleRed} or~\ref{sec:PoleRedAn} to
the polynomial in the numerator
$M^{
      N(\Gamma,\epsilon,\kappa)}(\underline x)+q_{N(\Gamma,\epsilon,\kappa)-1}(t,\epsilon)
    P^{N(\Gamma,\epsilon,\kappa)-1}(\underline x)$.
The resolution of the
system~\eqref{e:sysCFU} determines the  rational coefficient
$q_{N(\Gamma,\epsilon,\kappa)-1}(t,\epsilon)$ and
$M^{N(\Gamma,\epsilon,\kappa)-1}(x)$ computed using~(\ref{e:MfinalAnReg}). One
iterates the reduction until the power of  the second Symanzik
polynomial $\textbf{F}_\Gamma$ is $n-L-1$ so
that
\begin{equation}\label{e:stepfinal}
\left[\left(d\over dt\right)^{N(\Gamma,\epsilon,\kappa)}
  +\sum_{\mathsf{a}=1}^{N(\Gamma,\epsilon,\kappa)-1}
  {c_{N(\Gamma,\epsilon,\kappa)-\mathsf{a}}(t,\epsilon)\over c_{N(\Gamma,\epsilon,\kappa)}(t,\epsilon)} \left(d\over dt\right)^{N(\Gamma,\epsilon,\kappa)-\mathsf{a}}\right] \Omega_\Gamma^{\epsilon,\kappa}=M^{0}\,\Omega_\Gamma^{\epsilon,\kappa}+ d\beta_\Gamma^{\epsilon ,\kappa} \, ,
\end{equation}
where $M^0$ is of degree 0 so that
\begin{equation}
  q_0(t,\epsilon,\kappa)=  {c_{0}(t,\epsilon,\kappa)\over
    c_{N(\Gamma,\epsilon,\kappa)}(t,\epsilon,\kappa)}= -M^0  .
\end{equation}
The inhomogeneous term  $\beta_\Gamma^{\epsilon,\kappa} $ is the sum of 
$\beta^{\mathsf{a}}$ with $1\leq
\mathsf{a}\leq N(\Gamma,\epsilon,\kappa)$
contributions with their multiplicative factor as given
in~\eqref{e:Pred}
\begin{equation}
  \label{e:betaGamma}
  \beta_\Gamma^{\epsilon,\kappa}= \sum_{1\leq i<j\leq n} (x_i B^j_\Gamma- x_j B^i_\Gamma) \,
  dx_1\wedge \cdots \wedge \widehat{dx_i}\wedge \cdots \wedge\widehat{dx_j}\wedge
  \cdots \wedge dx_n 
\end{equation}
with
\begin{equation}
  \label{e:BetaGamma}
  \vec B^{\epsilon,\kappa}_\Gamma:=\left(\sum_{\mathsf{a}=1}^{N(\Gamma,\epsilon,\kappa)} {
      \vec C_{\mathsf{a}} (\underline x) \over
   ( \mathsf{a}-1+n-L(\delta-\epsilon))\textbf{F}_\Gamma^{\mathsf{a}-1}}\right) \,\omega_\Gamma^{\epsilon,\kappa} ,
\end{equation}
which is a vector of degree of homogeneity $1-n$ in the edge variables
$\underline x$.
Since $d\beta_\Gamma^{\epsilon,\kappa}= -\vec\nabla\cdot \vec B_\Gamma\,
\Omega_0^{(n)}$, from~\eqref{e:stepfinal} we have the ordinary
differential equation satisfied by the integrand of the Feynman
integral
\begin{equation}\label{e:stepfinalode}
\left[\left(d\over dt\right)^{N(\Gamma,\epsilon,\kappa)}
  +\sum_{r=1}^{N(\Gamma,\epsilon,\kappa)}
  {c_{N(\Gamma,\epsilon,\kappa)-r}(t,\epsilon)\over c_{N(\Gamma,\epsilon,\kappa)}(t,\epsilon)} \left(d\over dt\right)^{N(\Gamma,\epsilon,\kappa)-r}\right] \omega_\Gamma^{\epsilon,\kappa} =-\vec\nabla\cdot \vec B^{\epsilon,\kappa}_\Gamma\,.
\end{equation}
Integrating this expression over the positive orthant leads to
  the inhomogeneous differential equation satisfied by the Feynman
  integrals
  \begin{multline}\label{e:stepfinalodeInt}
\left[\left(d\over dt\right)^{N(\Gamma,\epsilon,\kappa)}
  +\sum_{r=1}^{N(\Gamma,\epsilon,\kappa)}
  {c_{N(\Gamma,\epsilon,\kappa)-r}(t,\epsilon)\over
    c_{N(\Gamma,\epsilon,\kappa)}(t,\epsilon)} \left(d\over
    dt\right)^{N(\Gamma,\epsilon,\kappa)-r}\right]
I_\Gamma^{\epsilon,\kappa} \cr=-\int_{x_i\geq0}\vec\nabla\cdot \vec
B^{\epsilon,\kappa}_\Gamma dx_1\cdots dx_n\,.
\end{multline}
The inhomogeneous term is a total derivative reflecting the fact that
it is spanned by Feynman integral with collapsed edge of the original
graph $\Gamma$. The evaluation of this inhomogeneous term is delicate
and requires taking into account the various blow-ups of the domain of integration so the integral is well-defined~\cite{bek,Brown:2009ta,Bloch:2016izu,muller2014picard}.

\medskip
 
 Let us comment on how to determine the order of the differential operators. An upper bound on the order can be computed
 from the number of master integrals.
 The number of master
integrals  can be computed from the Euler
characteristic of complement of the graph polynomial~\cite{Bitoun:2017nre} or counting the
critical points of the Euler representation of the integral in
projective
space~\cite{Lee:2013hzt,Cacciatori:2021nli}
or~\cite{Mastrolia:2018uzb,Frellesvig:2019uqt} (see~\cite{Agostini:2022cgv} for
a discussion of the relation between the different ways of computing the number of master
integrals). On the other hand, determining the minimal order is a
difficult question. Our pragmatical approach is to increase the order
starting from 
lower orders until the system~\eqref{e:sysCFU} has a solution, which is
the spirit of the Griffiths-Dwork reduction applied to Feynman
integrals in~\cite{Muller-Stach:2011qkg}.

In practice, for determining the minimal order it is enough to run the step of the
Griffiths-Dwork reduction for fixed generic numerical values for the physical
parameters (the internal masses and kinematic
parameters), because all the reduction amounts to solve a linear system in the projective space of
the edge parameters $\underline x$. This allows to determine the smallest order for which the
algorithm closes to give a differential operator.
The Griffiths-Dwork algorithm does not automatically
lead to an irreducible differential operator.
The factorisation of a linear ordinary
differential operator can be done with the {\tt DFactor} routine from
{\tt Maple} up to the order 4 for generic parameters~\cite{PutSinger,vanHoeij}, and to any
orders for linear  ordinary
differential operators with numerical coefficients
using the {\tt facto} algorithm in {\tt
	sagemath}~\cite{chyzak2022symbolic,goyer2021sage}.

 In the rational case $\epsilon=\kappa=0$, it was noticed
in~\cite{Bloch:2013tra,Bloch:2016izu,Bloch:2014qca,Lairez:2022zkj} that the order of the minimal differential operator is
smaller than the number of irreducible master integrals.
When the regulator take integer values the minimal order is smaller
than the number of master integrals. 
In the various cases studied below,
we will see that the order of the minimal differential operator can saturate
the upper bound given by the number of masters for generic values of
the regulators  $\epsilon$  and $\kappa$. 
This will
      be discussed further in Section~\ref{sec:minim-order-diff}.

\section{One-loop examples}\label{sec:oneloop}
\begin{figure}[ht]
	\centering
	\begin{tikzpicture}
		\draw[dashed](-1,-1)--(-1.3,-1.3);
		\draw[dashed](1,-1)--(1.3,-1.3);
		\draw[dashed](1,1)--(1.3,1.3);
		\draw[dashed](-1,1)--(-1.3,1.3);
		\draw[dashed, thick](-1,-1)--(-1.0,1.0)--(1.0,1.0)--(1.0,-1.0)--cycle;
		\node[text width=0.5cm, text centered ] at (-1.2,0) {$1$};
		\node[text width=0.5cm, text centered ] at (0,-1.3) {$2$};
		\node[text width=0.5cm, text centered ] at (1.2, 0) {$3$};
		\node[text width=0.5cm, text centered ] at (0,1.3) {$4$};
		\node[text width=0.5cm, text centered ] at (-1.5,1.5){$k_1$};
		\node[text width=0.5cm, text centered ] at (1.5,1.5){$k_2$};
		\node[text width=0.5cm, text centered ] at (1.5,-1.5){$k_3$};
		\node[text width=0.5cm, text centered ] at (-1.5,-1.5){$k_4$};
	\end{tikzpicture} 
\caption{The box  graph with massless external and internal states.
  The outgoing external momenta are $k_i$ with $k_1+\cdots+k_4=0$. The
          labels of the graph give the index of the edge variable $x_i$. }\label{fig:box}
\end{figure}
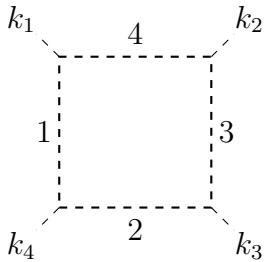

We start in Section~\ref{sec:box} with the simple example of the
massless  one-loop box of Fig.~\ref{fig:box} in dimensional regularisation, which will serve
as an illustration of where the main features of the algorithm emerge.

We then work out a Gr\"obner basis of differential operators
associated with the Cross Witten diagram for conformally coupled 
$\phi^4$  in  four dimensional  de Sitter space. It was shown
 in~\cite{Heckelbacher:2022fbx,Heckelbacher:2022hbq}  that the
 cosmological correlators of conformally coupled $\phi^4$ can be organised  as dimensionally regulated flat space Feynman integrals in position
space. Because of the measure of integration in (anti)-de Sitter the
resulting integrals fall into the category of the analytic
regularisation of Section~\ref{sec:Twisted}.

\subsection{The massless box graph in dimensional regularisation}\label{sec:box}
 We define the usual Mandelstam invariants as $t=(k_1+k_3)^2$, $s=(k_1+k_2)^2$.  The graph polynomials are given by
\begin{equation}
	\label{e:Boxgraphpolynomials}
	\textbf{U}_{\Box}=x_1+\cdots +x_4,\qquad
	\textbf{F}_{\Box}(s,t)=-t x_2x_4-sx_1x_3 \, .
\end{equation}
The twisted differential in Eq.~\eqref{e:OmegaTwistDimReg} for the box
graph in $D=4-2\epsilon$ in the projective
space $\mathbb P^3$ reads $\Omega_\Box^\epsilon= \omega_\Box^\epsilon
\,\Omega_0^{(4)}$, where
\begin{equation}\label{e:OmegaBox}
	\omega_\Box^\epsilon=   {1\over \textbf{F}_{\Box}(s,t)^2
	}\left(\textbf{U}_{\Box}^2\over \textbf{F}_{\Box}(s,t)\right)^\epsilon \, .
\end{equation}
This is a single scale function depending only on the ratio
$X=t/s$ of the kinematic invariants, so we scale the integral obtaining
$\tilde \Omega_\Box^\epsilon(X)=(-s)^{2+\epsilon} \Omega_\Box^\epsilon(s,X s)$.
The application of the procedure given in
Section~\ref{sec:deriv-diff-equat} needs only to start at the first
order. Computing the derivative with respect to $X$ we obtain 
\begin{equation}
P  ^{(1)}= (2+\epsilon) x_2 x_4 \, , 
\end{equation}
which we will reduce with respect to
\begin{equation}\vec\nabla(\textbf F_\Box)=\left(
    \partial_{x_1}\textbf{F}_\Box=-x_3, \partial_{x_2}\textbf{F}_\Box=
    -x_4 X, \partial_{x_3}\textbf{F}_\Box=-x_1, -X x_2  \right).
\end{equation}
The  vector  $\vec C_{1}(\underline x)$ has components 
\begin{equation}
C^i_{1}(\underline x)= \sum_{e \in m_{2,4}} \lambda^i_e x^e,	
\end{equation}
where $m_{i,j}$ denote the set of exponent vectors of degree $i$ in
$j$ variables.
Therefore, $m_{2,4}=\{(a_1,a_2,a_3,a_4)| a_1+a_2+a_3+a_4=2, a_1,a_2,a_3,a_4\geq0 \}$.
Since $\deg(C^i_{1}(\underline x))=2$, $\deg (c_1(\underline x))=1$,
i.e., $c_1(\underline x)=\sum_{e\in m_{1,4}} q_e x^e$, and
  $m_{1,4}=\{(1,0,0,0), (0,1,0,0), (0,0,1,0),\\ (0,0,0,1)\}$. Hence the linear system becomes
\begin{equation}\label{e:sysCFU-Box}
	\left\{\begin{array}{@{}l@{}}
\sum_{i=1}^4 \sum_{e \in m_{2,4}} \lambda^i_e x^e \partial_{x_i}(\textbf F_\Box) 
		=    (2+\epsilon) x_2 x_4\\
\sum_{i=1}^4\sum_{e \in m_{2,4}} \lambda^i_e x^e \partial_{x_i}(\textbf U_\Box) = \sum_{e\in m_{1,4}} q_e x^e \textbf{U}_\Box 
	\end{array}\right.\,,
\end{equation}
which leads to 
\begin{equation}
	c_0= \frac{1+X+\epsilon }{X(X+1)}.
\end{equation}
We then derive
\begin{equation}\label{e:PFBox}
	\mathscr{L}_\Box^\epsilon= (X+1) X {d\over dX}+(1+X+\epsilon)  \,.
\end{equation}
In this simple case the algorithm  requires only one iteration.
The boundary contribution in Eq.~\eqref{e:BetaGamma} is given by
\begin{align}
  \vec B_\Box^\epsilon&=\left\{X\frac{x_4}{2},-\frac{x_1+x_2+x_3}{2
                   },X\frac{x_4}{2},-\frac{x_4 (2 X+1)}{2
                   }\right\}\cr
                   &+\lambda_{0,0,1,0}^3 \left\{x_4
                     X-x_1,x_1-x_3,x_3-x_4 X,0\right\}\cr
                     &+\lambda_{0,0,1,0}^4 \left\{x_4-x_2,x_2 X-x_3,0,x_3-x_4 X\right\}\cr
                       &+\lambda_{1,0,0,0}^4 \left\{0,x_2 X-x_1,(x_4-x_2) X,x_1-x_4 X\right\}.
\end{align}
The boundary vector depends on the free coefficients $\lambda^i_{n_1,n_2,n_3,n_4}$ from the
reduction.  This freedom arises from the kernel of the linear
  system~\eqref{e:sysCFU-Box} on the unknown coefficients $\lambda^i_{n_1,\dots,n_4}$.
The gradient of this vector does not depend on the free coefficients
as it reads
\begin{equation}
  \vec\nabla\cdot\vec B_\Box^\epsilon=\frac{x_2 x_4 X (X \epsilon +X+1)-x_1 x_3 (X+\epsilon +1)}{X (X+1) (x_2 x_4 X+x_1
   x_3)}\, \omega_\Box^\epsilon  ,
\end{equation}
so that 
\begin{equation}
  	\mathscr{L}_\Box^\epsilon    \omega_\Box^\epsilon+\vec\nabla\cdot\vec B_\Box^\epsilon=0\,.
\end{equation}

By integrating  $\vec\nabla\cdot\vec B_\Box^\epsilon$ over the
$\Delta_4=\{x_i\geq0,1\leq i\leq 4\}$ we get the
inhomogeneous term 
\begin{equation}\label{e:SourceBox}
	\mathscr{S}_\Box^\epsilon=\frac{(\epsilon +1)\Gamma (-\epsilon -1)^2 }{\Gamma
		(-2 \epsilon )}  \, \left( (-s)^{-1-\epsilon }+(-t)^{-\epsilon -1}\right) \, .
\end{equation}
This simple example illustrates the general procedure as we will see with more loops. In general the system of equations is dense.

\subsection{The Cross Witten diagram in AdS$_4$ in dimensional regularisation}\label{sec:WittenCross}
\begin{figure}[ht]
	\centering
	\begin{tikzpicture}[very thick]
		\path [draw](-1,-1) --(1,1);
		\path [draw](-1,1) --(1,-1);
		\draw[blue] circle (1.4cm);
		\coordinate (A) at (0,0);   \filldraw (A) circle (2.pt);
		\coordinate (B) at (-1,1);   \filldraw (B) circle (2.pt);
		\coordinate (C) at (-1,-1);   \filldraw (C) circle (2.pt);
		\coordinate (D) at (1,1);   \filldraw (D) circle (2.pt);
		\coordinate (E) at (1,-1);   \filldraw (E) circle (2.pt);
		\node[text width=0.5cm, text centered ] at
                (-1.3,1.3){$\vec v_2$};
		\node[text width=0.5cm, text centered ] at
                (-1.3,-1.3){$\vec v_1$};
		\node[text width=0.5cm, text centered ] at
                (1.3,-1.3){$\vec v_4$};
		\node[text width=0.5cm, text centered ] at
                (1.3,1.3){$\vec v_3$};
	\end{tikzpicture}
  \caption{Witten Cross diagram, between four states on
	boundary of $dS_4$. }
\label{fig:cross}
	\end{figure}
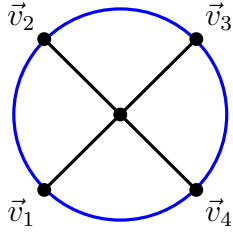
The dimensionally regulated Witten cross diagram of Fig.~\ref{fig:cross} considered in
Section~4.1 of~\cite{Heckelbacher:2022fbx} is given by the integral
over the bulk point $X$

\begin{equation}
       \mathcal{ W}_0^{1,4-4\varepsilon}(\zeta,\bar\zeta)=\frac12\frac{\zeta\bar\zeta}{(v_{12}v_{34})^2}\int_{\mathbb{R}^4}\frac{d^{4-4\varepsilon}X}{\|X\|^2\|X-u_1\|^{2(1-4\varepsilon)}\|X-u_{\zeta}\|^2}\,.
\end{equation}
with the vectors  $u_1=(1,0,0,0)$ and 
$u_{\zeta}=\left(\frac{\zeta+\bar\zeta}{2},\frac{\zeta-\bar\zeta}{2i},0,0\right)$ where $\zeta$  and $\bar\zeta$ parameterise the cross ratios $\zeta\bar\zeta=v_{12}^2v_{34}^2/(v_{14}^2v_{23}^2)$ and
  $(1-\zeta)(1-\bar\zeta)=v_{13}^2v_{24}^2 /(v_{14}^2 v_{23}^2)$
  defined by the position on the boundary of (A)dS$_4$.

The parametric representation is given by
$
  I^\varepsilon_\times(\zeta,\bar\zeta)= \int_{\Delta_3}   \Omega_\times^{2\varepsilon,-4\varepsilon}(\zeta,\bar\zeta)
$ the integration over the twisted differential form
\begin{equation}
  \Omega_\times^{2\varepsilon,2\varepsilon}(\zeta,\bar\zeta)={\pi^{2-2\varepsilon}\Gamma(1-2\varepsilon)\over\Gamma(1-4\varepsilon)} {1\over \textbf{U}_\times
    \textbf{F}_\times (\zeta,\bar\zeta)} \left(
    \textbf{F}_\times (\zeta,\bar\zeta)^{2}\over  x_2^{4}\right)^\varepsilon  \, \Omega_0^{(3)}
\end{equation}
over the domain $\Delta_3=\{x_i\geq0, 1\leq i\leq 3\}$ and with the graph polynomials
\begin{equation}
  \textbf{U}_\times= x_1+x_2+x_3,  \qquad \textbf{F}_\times (\zeta,\bar\zeta)= x_1x_2+
  \zeta\bar\zeta x_1x_3+ (1-\zeta)(1-\bar\zeta)x_2x_3  .
\end{equation}
We can apply the algorithm to the case of the analytic regularisation
of Section~\ref{sec:PoleRedAn} 
with $\delta=2$, $\epsilon=\kappa=2\varepsilon$, $(p_1,p_2,p_3)=(0,-2,0)$. 
We find the following set of  differential operators
acting on $I_\times(\zeta,\bar\zeta)$
\begin{multline}
  \mathscr{L}_{\times,1}= (\zeta -1) (\zeta -\bar\zeta) \zeta ^2{\partial^2\over\partial\zeta^2}  +
   \left(\zeta (3 \zeta -\bar\zeta-2)-2\varepsilon  \left(\zeta ^2+\zeta 
   \bar\zeta-2 \bar\zeta\right) \right) \zeta {\partial\over\partial\zeta}\cr+(2 \varepsilon -1) \left(2 \varepsilon  (\zeta +\bar\zeta)-\zeta ^2\right)
\end{multline}
and
\begin{equation}
  \mathscr{L}_{\times,2}= (\zeta -1)\zeta\bar\zeta{\partial\over\partial\zeta}  +
   (\bar\zeta -1)\zeta\bar\zeta{\partial\over\partial\bar\zeta}+\zeta\bar\zeta-2\varepsilon(\zeta+\bar\zeta).
 \end{equation}
We have checked  that
they form a Gr\"obner basis with respect to the lexicographical
ordering of $\zeta,\bar\zeta$ using 
the command {\tt OreGroebnerBasis}, which is part of the package {\tt
	HolonomicFunctions}~\cite{Koutchan}. 

 The algorithm determines the boundary terms 
 such that\footnote{Details are provided in the  attached {\tt
     Mathematica}  notebook with  this example is accessible at \href{https://github.com/pierrevanhove/TwistedGriffithsDwork/blob/main/Mathematica/Cross-AdS.pdf}{Cross-AdS.pdf}.}
 \begin{equation}
   \mathscr{L}_{\times,r}    \left( {1\over \textbf{U}_\times
    \textbf{F}_\times (\zeta,\bar\zeta)} \left(
    \textbf{F}_\times (\zeta,\bar\zeta)^{2}\over
    x_2^{4}\right)^\varepsilon \right)+ \vec\nabla\cdot\vec B^{2\varepsilon,2\varepsilon}_{\times,r}=0.
\qquad r=1,2.
\end{equation}
For computing the boundary contribution to say, $
\mathscr{S}_{\times,1}$ we need to evaluate
\begin{equation}
    \mathscr{S}_{\times,1}=- \int_{\Delta_3} \vec\nabla\cdot\vec
    B^{2\varepsilon,2\varepsilon}_{\times,1}\,.
\end{equation}
Integrating by parts we have the expression
\begin{multline}
    \mathscr{S}_{\times,1}=\iint_0^\infty
\left(\lim_{x_1\to\infty}
  (B^{2\varepsilon,2\varepsilon}_{\times,1})_1-\lim_{x_1\to0}
  (B^{2\varepsilon,2\varepsilon}_{\times,1})_1\right)dx_2dx_3\cr
+\iint_0^\infty
\left(\lim_{x_2\to\infty}
  (B^{2\varepsilon,2\varepsilon}_{\times,1})_2-\lim_{x_2\to0}
  (B^{2\varepsilon,2\varepsilon}_{\times,1})_2\right)dx_1dx_3
+\cr\iint_0^\infty
\left(\lim_{x_3\to\infty}
  (B^{2\varepsilon,2\varepsilon}_{\times,1})_3-\lim_{x_3\to0}
  (B^{2\varepsilon,2\varepsilon}_{\times,1})_3\right)dx_1dx_2\,.
\end{multline}
The components of $\vec B^{2\varepsilon,2\varepsilon}_{\times,1}$
depend on seven free parameters
\begin{equation}
  \label{e:varBcross}
  \left\{\lambda^1_{1,0,0},\lambda^3_{1,1,1},\lambda^3_{1,2,0},\lambda^3_{2,1,0},\lambda^3_{0,0,1},\lambda^3_{0,1,0},\lambda^3_{1,0,0}\right\}.
\end{equation}
For all values of these parameters we have for $\varepsilon>0$
\begin{equation}
    \lim_{x_2\to\infty}
  (B^{2\varepsilon,2\varepsilon}_{\times,1})_2=\lim_{x_2\to0}
  (B^{2\varepsilon,2\varepsilon}_{\times,1})_2=0\,.
\end{equation}
For all values of the parameters, the limits 
$\lim_{x_1\to0}(B^{2\varepsilon,2\varepsilon}_{\times,1})_1$ and $\lim_{x_3\to0}(B^{2\varepsilon,2\varepsilon}_{\times,1})_3$ 
are finite for $\varepsilon>0$.
But the limits $\lim_{x_1\to\infty}(B^{2\varepsilon,2\varepsilon}_{\times,1})_1$ and $\lim_{x_3\to\infty}(B^{2\varepsilon,2\varepsilon}_{\times,1})_3$ 
are not finite for $\varepsilon>0$.
With the choice of the parameters 
\begin{align}
 \lambda^3_{1,0,0}&= -\frac{4 \varepsilon -1 }{2 (\varepsilon
                       -1)} \lambda^3_{2,1,0},\cr
                       \lambda^3_{0,1,0}&=\frac{(2 \varepsilon -1) \lambda^3_{1,1,1}}{\zeta  \bar\zeta (\varepsilon
   -1)}+\frac{2 (2 \varepsilon -1) \lambda^3_{0,0,1}}{\zeta  \bar\zeta (4 \varepsilon
                                          -1)}-\frac{(2 \varepsilon -1)
   \lambda^3_{1,2,0}}{\varepsilon -1}\cr
                                          &-\frac{\lambda^3_{2,1,0} (2 \zeta +2 \bar\zeta-\zeta  \bar\zeta -2+4( \zeta  \bar\zeta - \zeta  - \bar\zeta 
                                            +1 )\varepsilon)}{2 \zeta  \bar\zeta (\varepsilon -1)}\cr
                                            &+\frac{(2 \varepsilon -1)^2 \left(\zeta
   +\bar\zeta -3 \zeta ^2+2 (4\zeta ^2  - \zeta   - \bar\zeta) \varepsilon
   \right)}{ (4 \varepsilon -1)(\zeta -1) \zeta ^3 \bar\zeta (\zeta -\bar\zeta)}\, , 
\end{align}
we have
$\lim_{x_1\to\infty}(B^{2\varepsilon,2\varepsilon}_{\times,1})_1=\lim_{x_3\to\infty}(B^{2\varepsilon,2\varepsilon}_{\times,1})_3=0$
for $\varepsilon>0$.  The boundary term is given by
  \begin{align}
    \mathscr{S}_{\times,1}&=-\iint_0^\infty
\lim_{x_1\to0}
  (B^{2\varepsilon,2\varepsilon}_{\times,1})_1\,dx_2dx_3
-\iint_0^\infty
\lim_{x_3\to0}
                            (B^{2\varepsilon,2\varepsilon}_{\times,1})_3\,dx_1dx_2,\cr
                            &=-\frac{\pi  (2 \varepsilon -1) \left((\zeta -1)^{2 \varepsilon }
   (\bar\zeta-1)^{2 \varepsilon }-1\right)}{ \sin (2 \pi  \varepsilon ) }\,.
\end{align}  
It was shown in~\cite{Heckelbacher:2022fbx} that
\begin{equation}
   I^\varepsilon_\times(\zeta,\bar\zeta)=
   2{\Li_2(\zeta)-\Li_2(\bar\zeta)-\frac12\log(\zeta\bar\zeta)\log\left(1-\zeta\over
    1-\bar\zeta \right) \over \zeta-\bar\zeta}+O(\varepsilon)\,,
\end{equation}
where $\Li_2(z)=-\int_0^z \log(1-u)d\log u$ is the dilogarithm.
One easily checks that
\begin{equation}
  \mathscr{L}_{\times,1}     I^\varepsilon_\times(\zeta,\bar\zeta)+ \mathscr{S}_{\times,1} =0+O(\varepsilon)\,.
\end{equation}
This analysis shows that the boundary term produced by the
Griffiths-Dwork reduction is not guaranteed to be integrable over the
positive orthant. The choice of free parameters we have made is
equivalent to add a total derivative (or an exact form) to get a
convergent integral. Although the
	algorithm produces the integrand for computing it according~\eqref{e:stepfinalodeInt}, this  simple case shows that evaluating the
inhomogeneous term is not an easy task, which will not carry for most
of the remaining cases studied in this paper.

\section{Two-loop examples}\label{sec:twoloop}
In this section we apply the algorithm to  the case of dimensionally
 regulated two loop integrals. We derive the $\epsilon$-deformation of
the equal-mass sunset integral in Section~\ref{sec:2sunset1mass}, the general mass configuration
sunset integral in Section~\ref{sec:2sunset3mass}, the two point kite
integral in Section~\ref{sec:kite}, the ice-cream cone
graph in Section~\ref{sec:ice-cream}, the non-planar  triangle-box  
graph in Section~\ref{sec:chicken} and the four points planar
and non-planar boxes~\ref{sec:doubleboxes}. We conclude in  Section~\ref{sec:Wittenicecream}
with the ice-cream cone graph in  analytic regularisation in four
dimensions, which arises in the two-loop correction to cosmological
correlators of  conformally coupled $\phi^4$  in  de
 Sitter space~\cite{Chowdhury:2023arc}.
\subsection{The two-point  two-loop sunset graph}
\label{sec:two-loop-case}
	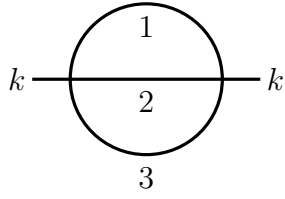
\begin{figure}[ht]
		\centering
		\begin{tikzpicture}
			\draw [very thick] (0,1)circle (1);
			\draw[very thick](-1.5,1)--(1.5,1);
			\node[text width=0.5cm, text centered ] at (0,0.7) {$2$};
			\node[text width=0.5cm, text centered ] at (0,1.7) {$1$};
			\node[text width=0.5cm, text centered ] at (0,-0.3) {$3$};
			\node[text width=0.5cm, text centered ] at
                        (-1.7,1.0) {$k$};
                        	\node[text width=0.5cm, text centered ] at (1.7,1.0) {$k$};
		\end{tikzpicture} 
		\caption{The two-loop sunset graph.  The
          labels of the graph give the index of the edge variable $x_i$.}
		  \label{fig:sunset2loop}
	\end{figure}
We now turn to the two-loop sunset graph of Fig.~\ref{fig:sunset2loop}
and show how to adapt the
Griffiths-Dwork reduction used in~\cite{Bloch:2016izu,Lairez:2022zkj}
to the $\epsilon$-dependent integrand from dimensional regularisation. For the two-loop case the differential
Eq.~\eqref{e:OmegaTwistDimReg}, setting $t=k^2$, is
\begin{equation}\label{e:OmegaSunset}
  \Omega_{\su}^{\epsilon}(t)={1\over
    \textbf{F}_\su(t)}\left(\textbf{U}_\su^3\over \textbf{F}_\su(t)^2\right)^\epsilon  \,\Omega_0^{(3)}
\end{equation}
with the graph polynomials 
\begin{align}
 \textbf{U}_\su&=x_1x_2+x_1x_3+x_2x_3,\cr
  \textbf{F}_\su(t)&=-t x_1x_2x_3 + (m_1^2x_1+m_2^2x_2+m_3^2x_3) \textbf{U}_\su.
\end{align}
The twisted differential form $\Omega_\su^{\epsilon}$ is defined on the
complement of the sunset elliptic curve $\textbf{F}_\su(t)=0$.
We discuss the equal-mass and the general case separately.
We apply the algorithm by starting to seek an operator of second order which is enough for the equal-mass case, but for the
three-mass case we find that the minimal order is four, in
agreement with previous results~\cite{Caffo:1998du,Remiddi:2013joa,Adams:2013nia}.

\subsubsection{The equal-mass case}\label{sec:2sunset1mass}
We derive the differential  operator satisfied by the two-loop all
equal-mass in general dimensions. 
We  start with at $N(\su,\epsilon)=2$ finding
\begin{equation}
 \left(d\over dt\right)^2
 \Omega_{\su}^{\epsilon}(t,\epsilon)={\Gamma(3+2\epsilon)\over\Gamma(1+2\epsilon)} {(x_1x_2x_3)^2\over \textbf{F}_\su(t)^3}
 \left(\textbf{U}_\su^3\over \textbf{F}_\su(t)^2\right)^\epsilon  \Omega_0^{(3)} ,
\end{equation}
from which we can extract $P^{(2)}=2 (\epsilon +1) (2 \epsilon +1)(x_1x_2x_3)^2$. 
Accordingly, we label the unknowns with the upper index so $S^{(k)}$ indicates the $k$-th reduction. We then  perform the Jacobian reduction of $P^{(2)}$ as
\begin{equation}\label{e:reduc}
  2 (\epsilon +1) (2 \epsilon +1)\,  (x_1x_2x_3)^2= \sum_{i=1}^3 
  \sum_{e\in m_{4,3}}\lambda^{(2),i}_{e} x^e \partial_{x_i}  (\textbf{F}_\su(t)) \,,
\end{equation}
where we have written explicitly the components of $\vec{C}^{(2)}$ as $C^{(2)}_i=\sum_{e\in m_{4,3}}\lambda^{(2),i}_{e} x^e$.
We have to solve the following equation for the coefficients $\lambda_e^{(2),i}$ coupled to the equations generated by
\begin{equation}\label{e:C2red}
\sum_{i=1}^3 
\sum_{e\in m_{4,3}}\lambda^{(2),i}_{e} x^e \partial_{x_i}(\textbf U_\su) = \sum_{e\in m_{3,3}} q^{(2)}_e x^e \textbf{U}_\su \,, 
\end{equation}
where $c^{(2)}=\sum_{e\in m_{3,3}} q^{(2)}_e x^e$ is an unknown homogeneous degree 3 polynomial
in $\underline x$.
The unknowns will be fully determined at the end of  algorithm. For this case, 
the differential form given in Eq.~\eqref{e:betadef} reads
 \begin{multline}
 	\label{eq:beta-equalmass-sunset}
   \beta_\su^\epsilon=   {x_2  C_3^{(2)}-x_3  C_2^{(2)}\over
     \textbf{F}_\su(t)^2} \left(\textbf{U}_\su^3\over \textbf{F}_\su(t)^2\right)^\epsilon dx_1+   {x_3  C_1^{(2)}-x_1 C_3^{(2)}\over
     \textbf{F}_\su(t)^2} \left(\textbf{U}_\su^3\over \textbf{F}_\su(t)^2\right)^\epsilon dx_2\cr
   +  {x_1  C_2^{(2)}-x_2  C_1^{(2)}\over
    \textbf{F}_\su(t)^2} \left(\textbf{U}_\su^3\over \textbf{F}_\su(t)^2\right)^\epsilon dx_3, 
 \end{multline}
which leads to 
\begin{multline}
 (2+2\epsilon)\left(d\over dt\right)^2
  \Omega_{\su}^{\epsilon}(t,\epsilon)=
 \left({\sum_{i=1}^3 \partial_i C_i^{(2)}\over
     \textbf{F}_\su(t)^{2}}+3\epsilon{\sum_{i=1}^3 C_i^{(2)}\partial_i
     \log \textbf{U}_\su\over \textbf{F}_\su(t)^{2}}\right)\cr \times\left(\textbf{U}_\su^3\over \textbf{F}_\su(t)^2\right)^\epsilon \Omega_0^{(3)}-d(2\beta_\su^\epsilon).
\end{multline}
Therefore, using Eq.\eqref{e:Mdimregfinal}, we define
\begin{equation}
  \label{e:M2}
  M^{(2)}:={\sum_{i=1}^3 \partial^i  C_i^{(2)}+3\epsilon
    c^{(2)}\over 2+2\epsilon} \,, 
\end{equation}
which upon using~\eqref{e:C2red} leads
 to the reduction of the pole
\begin{equation}
 \left(d\over dt\right)^2
  \Omega_{\su}^{\epsilon}(t,\epsilon)=
{M^{(2)}\over
     \textbf{F}_\su^{2}}\left(\textbf{U}_\su^3\over \textbf{F}_\su(t)^2\right)^\epsilon \Omega_0^{(3)}+d\beta_\su^\epsilon \, .
 \end{equation}
We now add the first derivative with an unknown rational coefficient $q_1(t,\epsilon)$
\begin{multline}
 \left(d\over dt\right)^2
  \Omega_{\su}^{\epsilon}(t,\epsilon) +q_1(t,\epsilon) \left(d\over dt\right)
  \Omega_{\su}^{\epsilon}(t,\epsilon) =
{M^{(1)}\over
     \textbf{F}_\su(t)^{2}}\left(\textbf{U}_\su^3\over \textbf{F}_\su(t)^2\right)^\epsilon \Omega_0^{(3)}+d\beta_\su^\epsilon \, , 
 \end{multline}
where $M^{(1)}:=M^{(2)}+q_1(t,\epsilon) x_1x_2x_3(1+2\epsilon)$.
 We then reduce the pole a second time by writing
 \begin{equation}\label{e:M2red}
   M^{(1)}= \sum_{i=1}^3
   C_i^{(1)} \partial^i \textbf{F}_\su(t) \,, 
 \end{equation}
 where $ C_i^{(1)}$ are unknown homogeneous degree 1 polynomials.
Thus,
\begin{equation}
 \left(d\over dt\right)^2
  \Omega_{\su}^{\epsilon}(t,\epsilon) +q_1(t,\epsilon) \left(d\over dt\right)
  \Omega_{\su}^{\epsilon}(t,\epsilon) =
{ \sum_{i=1}^3
   C_i^{(1)} \partial^i \textbf{F}_\su(t)\over
     \textbf{F}_\su(t)^{2}}\left(\textbf{U}_\su^3\over \textbf{F}_\su(t)^2\right)^\epsilon \Omega_0^{(3)}+d\beta_\su^\epsilon \,.
 \end{equation}
The last step to compute the differential operator is to derive the constant term, which must reduce the poler order of the right-hand-side of this equation. We thus impose
\begin{equation}\label{e:c0}
  \sum_{i=1}^3C_i^{(1)}\partial^i \textbf{F}_\su(t)+ q_0(t,\epsilon) \textbf{F}_\su (t) =0.
\end{equation}
Solving all the equations needed for the pole reduction leads
to the unique solution for the coefficients $q_1$ and $q_0$. 
The solutions are 
\begin{align}
  q_1(t,\epsilon)&=\frac{\left(3 t^2-10 t-9\right) \epsilon }{(t-9) (t-1) t}+\frac{3 t^2-20 t+9}{(t-9) (t-1)
   t}\, ,\cr
  q_0(t,\epsilon)&=\frac{\epsilon ^2 (2 t+2)}{(t-9) (t-1) t}+\frac{\epsilon  (3 t-5)}{(t-9) (t-1)
   t}+\frac{t-3}{(t-9) (t-1) t} \, ,
\end{align}
leading to the $\epsilon$-deformed differential operator
\begin{multline}
  \label{e:PF2sunset1massepsilon}
     \mathscr{L}^{(2),\epsilon,1-mass}_\su ={d\over dt}\left( t(t-1)(t-9)
       {d\over dt}\right)+(t-3)\cr+\epsilon\left((3t^2-10t-9) {d\over
         dt}+3t-5\right)
     +\epsilon^2 2(t+1).
\end{multline}
Collecting the inhomogeneous contributions into the vector
\begin{equation}
  \vec B_\su^\epsilon=\sum_{\mathsf{a}=1}^2 {\vec C_{\mathsf{a}}\over
    (\mathsf{a}+2\epsilon)\textbf{F}_\su(t)^{\mathsf{a}-1}}\, \Omega_\su^\epsilon
\end{equation}
one can check that the action of this differential operator on
$\Omega_{\su}^\epsilon(t)$ is
\begin{equation}
      \mathscr{L}^{(2),\epsilon,1-mass}_\su
      \Omega_{\su}^\epsilon(t)+\vec\nabla \cdot\vec B_\su^\epsilon=0
    \end{equation}
as it should be from the general considerations of Section~\ref{sec:deriv-diff-equat}.

\medskip
From the solutions, we can compute the  inhomogeneous term is given by evaluating the integral 
\begin{equation}
  \mathscr{S}_\su(t,\epsilon)
  =-\int_{\Delta_3} \vec\nabla\cdot\vec B^\epsilon_\su \,.
\end{equation}
Because the denominator of $ B^\epsilon_\su$ has a pole at the coordinate
point $[1:0:0]$, $[0:1:0]$ and $[0:0:1]$ one needs to consider the
blow-up  of the domain of integration $\Delta_3$. This is done by inserting a small
$\mathbb P^1$ of radius $\rho$ (see Eq.~(3.47) of~\cite{Bloch:2016izu})
\begin{equation}
    \mathscr{S}_\su(t,\epsilon)=\lim_{\rho\to0} \sum_{i=1}^3
    \int_{\partial\tilde\Delta_3|_{x_i=0}} \sum_{1\leq j\neq i\leq 3}
    (B^\epsilon_\su)_j dx_j \, .
\end{equation}
A computation identical to the one performed in~\cite{Bloch:2016izu} leads to
\begin{equation}
 \mathscr{S}_\su(t,\epsilon)=  -6 \, {\Gamma(1+\epsilon)^2\over \Gamma(1+2\epsilon)} \, . 
\end{equation}
The piece of order $\epsilon^0$ reproduces the differential operator
for the two-loop equal-mass sunset in $D=2$ given
in~\cite{Vanhove:2014wqa}. This  differential equation in
  general dimensions can be
  obtained by applying the results~\cite{Remiddi:2013joa,Remiddi:2016gno} to the all-equal-mass case.

\subsubsection{The different mass case}\label{sec:2sunset3mass}

For the non-equal-mass case the order of the differential equation is
4 with the following $\epsilon$ expansion
\begin{equation}
     \mathscr{L}^{\epsilon}_\su =   \mathscr{L}^{(1)}_1
     \mathscr{L}^{(2)}_1    \mathscr{L}^{3-mass}_\su +\epsilon
     \mathscr{L}^{(3)}_4+\epsilon^2  \mathscr{L}^{(4)}_3+\epsilon^3
     \mathscr{L}^{(5)}_2+ \epsilon^4 \mathscr{L}^{(6)}_1 +\epsilon^5
    \mathscr{L}^{(7)}_0,
   \end{equation} 
   where $ \mathscr{L}^{(r)}_i$  are irreducible differential operator
   of  order $i$ and $\mathscr{L}^{3-mass}_\su$ is the differential
   operator for the three-mass two-loop sunset integral in two
   dimensions. This differential equation reproduces the one 
   derived in~\cite{Remiddi:2013joa,Remiddi:2016gno}.
The $\epsilon$ deformation does not change the non-apparent
singularities of the differential operator as can be seen from the
coefficient of the highest order term
\begin{multline}
  \mathscr{L}^{\epsilon}_\su \Big\vert_{(d/dt)^4}= t^3\prod_{i=1}^4 (t-
  \mu_i^2) \Big(-\left(2 \epsilon +5\right) t^{2}-2
    \left(m_{1}^{2}+m_{2}^{2}+m_{3}^{2}\right) \left(1+2 \epsilon
    \right) t \cr+ \left(7+6 \epsilon \right)\prod_{i=1}^4 \mu_i
\Big)  \, , 
\end{multline}
where $\mu_i=\{m_1+m_2+m_3,-m_1+m_2+m_3,m_1-m_2+m_3,m_1+m_2-m_3\}$ are
the thresholds.  The $\epsilon$ deformation is only affecting the
apparent singularities, since the $\epsilon$ factor
in~\eqref{e:OmegaSunset} does not change the nature of the singular
locus which is still given by the same elliptic curve as in the
$\epsilon=0$ case.
  
The action  of $\mathscr{L}^{\epsilon}_\su$
on the Feynman integral is given by 
\begin{equation}
     \mathscr{L}^{\epsilon}_\su  I_\su^\epsilon(\underline
     m,t,\epsilon)=\mathscr{S}_\su(\vec m,t,\epsilon) 
   \end{equation}
   with the source term
   \begin{equation}
     \mathscr{S}_\su(\vec m,t,\epsilon)=\frac{c_{23}(t,\epsilon)\Gamma (\epsilon +1)^2}{ (m_{2} m_{3})^{2 \epsilon}\Gamma (1+2\epsilon)}+\frac{c_{13}(t,\epsilon)\Gamma (\epsilon +1)^2}{ (m_{1} m_{3})^{2 \epsilon }\Gamma (1+2
   \epsilon)}+\frac{c_{12}(t,\epsilon)\Gamma (\epsilon +1)^2}{ (m_{1} m_{2})^{2 \epsilon }\Gamma (1+2
   \epsilon )} \, , 
\end{equation}
where $c_{12}(t,\epsilon)$, $c_{13}(t,\epsilon)$ and
$c_{23}(t,\epsilon)$ are polynomials of degree 4  in $t$ and degree 2 in
$\epsilon$, respectively.
The contribution to the inhomogeneous term arise from each
  boundary contributions located at $x_1=0$, $x_2=0$ and $x_3=0$. They
  are given by the    two-bouquet Feynman graphs
  \begin{equation}
         \mathscr{S}_\su(\vec m,t,\epsilon)=
         \begin{gathered}
           \begin{tikzpicture}
\begin{feynman}
\vertex (a){\(k\)};
\vertex[below=2 cm of a](d){\(k\)};
\vertex[above=1cm of d](e);
\vertex[left=1cm of e,label={\(m_2~~~~\)}](h);
\vertex[right=1cm of e,label={\(~~~~m_3\)}](g);
\diagram*{
    (d)-- [very thick](a), (e)--[out=135,in=90,min distance=0.5cm,very thick](h), 
    (e)--[out=-135,in=-90,min distance=0.5cm,very thick](h),
    (e)--[out=45,in=90,min distance=0.5cm,very thick](g), 
    (e)--[out=-45,in=-90,min distance=0.5cm,very thick](g)
};
\end{feynman}
\end{tikzpicture}
\end{gathered}+
\begin{gathered}
           \begin{tikzpicture}
\begin{feynman}
\vertex (a){\(k\)};
\vertex[below=2 cm of a](d){\(k\)};
\vertex[above=1cm of d](e);
\vertex[left=1cm of e,label={\(m_1~~~~\)}](h);
\vertex[right=1cm of e,label={\(~~~~m_3\)}](g);
\diagram*{
    (d)-- [very thick](a), (e)--[out=135,in=90,min distance=0.5cm,very thick](h), 
    (e)--[out=-135,in=-90,min distance=0.5cm,very thick](h),
    (e)--[out=45,in=90,min distance=0.5cm,very thick](g), 
    (e)--[out=-45,in=-90,min distance=0.5cm,very thick](g)
};
\end{feynman}
\end{tikzpicture}
\end{gathered}+\begin{gathered}
           \begin{tikzpicture}
\begin{feynman}
\vertex (a){\(k\)};
\vertex[below=2 cm of a](d){\(k\)};
\vertex[above=1cm of d](e);
\vertex[left=1cm of e,label={\(m_1~~~~\)}](h);
\vertex[right=1cm of e,label={\(~~~~m_2\)}](g);
\diagram*{
    (d)-- [very thick](a), (e)--[out=135,in=90,min distance=0.5cm,very thick](h), 
    (e)--[out=-135,in=-90,min distance=0.5cm,very thick](h),
    (e)--[out=45,in=90,min distance=0.5cm,very thick](g), 
    (e)--[out=-45,in=-90,min distance=0.5cm,very thick](g)
};
\end{feynman}
\end{tikzpicture}
\end{gathered}    .   \end{equation}
      The coefficients match the one  derived from the general dimension
       results of~\cite{Remiddi:2013joa,Remiddi:2016gno}.

The results are provided on the {\tt
  SageMath} worksheet \href{https://nbviewer.org/github/pierrevanhove/TwistedGriffithsDwork/blob/main/Worksheets/Sunset-Twoloop-3mass-Epsilon.ipynb}{Sunset-Twoloop-3mass-Epsilon.ipynb}.
Expanding in powers of $\epsilon$, we have
\begin{equation}
    \mathscr{S}_\su(\vec m,t,\epsilon)=\mathscr{S}_\su^0(\vec m,t)
    +\left(c^{(1)}_0(\vec m)+
    \sum_{i=1}^3  c^{(1)}_i(\vec m)\log(m_i)\right)\,  \epsilon+O(\epsilon^2)
\end{equation}
with  the leading term given by
   \begin{equation}
   \mathscr{S}_\su^0(\vec m,t)=60 t^{4}+56\left( m_{1}^{2}+ m_{2}^{2}+
     m_{3}^{2}\right) t^{3}
   -308 \prod_{i=1}^4 \mu_i.
 \end{equation}
For $\epsilon=0$ the two-loop sunset integral satisfies  the
differential equation~\cite{Adams:2013nia,Bloch:2016izu}
\begin{equation}
  \mathscr{L}^{3-mass}_\su f_\su^{(0)}(t)= s_0(\vec m,t)+ \sum_{i=1}^3
  s_i(\vec m,t)\log(m_i^2)  
\end{equation}
with 
\begin{align}
  &s_0(\vec m,t)=\\
  &\quad 18 t^{4}-24\left(m_{1}^{2}m_{2}^{2}+
       m_{3}^{2}\right) t^{3}-4\left( m_{1}^{4}+m_2^4+m_3^4+10(
       m_{1}^{2} m_{2}^{2}+ m_{1}^{2}
       m_{3}^{2}+m_{2}^{2}
       m_{3}^{2})\right) t^{2}\nonumber\\
       &\quad+8
       \left(m_{1}^{2}+m_{2}^{2}+m_{3}^{2}\right)\prod_{i=1}^4\mu_i
       t +2 \prod_{i=1}^4\mu_i^2\,, \nonumber\\
  &s_1(\vec m,t)=\\
  &\quad\left(4 m_{1}^{2}-2 m_{2}^{2}-2 m_{3}^{2}\right)
       t^{3}+\left(-12 m_{1}^{4}+14 m_{1}^{2} m_{2}^{2}+14 m_{1}^{2}
       m_{3}^{2}+6 m_{2}^{4}-28 m_{2}^{2} m_{3}^{2}+6 m_{3}^{4}\right)
       t^{2}\nonumber\\
       &\quad+\Big(12 m_{1}^{6}-22 m_{1}^{4} m_{2}^{2}-22 m_{1}^{4} m_{3}^{2}+16 m_{1}^{2} m_{2}^{4}+16 m_{1}^{2} m_{3}^{4}-6 m_{2}^{6}+6 m_{2}^{4} m_{3}^{2}+6 m_{2}^{2} m_{3}^{4}\nonumber\\&\quad-6 m_{3}^{6}\Big) t-2 \left(2 m_{1}^{4}-m_{1}^{2} m_{2}^{2}-m_{1}^{2} m_{3}^{2}-m_{2}^{4}+2 m_{2}^{2} m_{3}^{2}-m_{3}^{4}\right) \prod_{i=1}^4\mu_i \,,
       \nonumber
       \end{align}
   \begin{align}
 &      s_2(\vec m,t)=\\
&\quad       \left(-2 m_{1}^{2}+4 m_{2}^{2}-2 m_{3}^{2}\right)
            t^{3}+\left(6 m_{1}^{4}+14 m_{1}^{2} m_{2}^{2}-28
            m_{1}^{2} m_{3}^{2}-12 m_{2}^{4}+14 m_{2}^{2} m_{3}^{2}+6
            m_{3}^{4}\right) t^{2}\nonumber\\
            &\quad     +\Big(-6 m_{1}^{6}+16 m_{1}^{4} m_{2}^{2}+6 m_{1}^{4} m_{3}^{2}-22 m_{1}^{2} m_{2}^{4}+6 m_{1}^{2} m_{3}^{4}+12 m_{2}^{6}-22 m_{2}^{4} m_{3}^{2}+16 m_{2}^{2} m_{3}^{4}\nonumber\\&\quad     -6 m_{3}^{6}\Big) t +2 \left(m_{1}^{4}+m_{1}^{2} m_{2}^{2}-2 m_{1}^{2} m_{3}^{2}-2 m_{2}^{4}+m_{2}^{2} m_{3}^{2}+m_{3}^{4}\right) \prod_{i=1}^4\mu_i
\,, \nonumber
            \end{align}
   with
   \begin{equation}
     s_3(\vec m,t)=-s_1-s_2\,.
   \end{equation}
It can be checked that 
\begin{equation}
  \mathscr{S}_\su^0(\vec m,t)=   \mathscr{L}^{(1)}_1
     \mathscr{L}^{(2)}_1    \mathscr{L}^{3-mass}_\su  f_\su^{(0)}(t).
\end{equation}
The logarithmic dependence on the masses arise at the order $\epsilon$ 
 and
 \begin{align}
&     c^{(1)}_0(\vec m)=114 t^{4}+168\left(m_{1}^{2}+ m_{2}^{2}+ m_{3}^{2}\right)
  t^{3}+\Big(-552 (m_{1}^{4}+m_2^4+m_3^4)\nonumber\\&
  \quad +1136 (m_{1}^{2} m_{2}^{2}+
    m_{1}^{2} m_{3}^{2}+ m_{2}^{2} m_{3}^{2})\Big) t^{2}-64
  \left(m_{1}^{2}+m_{2}^{2}+m_{3}^{2}\right) \prod_{i=1}^4\mu_i t +14
   \prod_{i=1}^4\mu_i^2\,, \end{align}
 and
 \begin{align}
& c^{(1)}_1(m_1,m_2,m_3)=\\
&\quad-80 t^{4}-\left( 88 m_{1}^{2}+ 68 m_{2}^{2}+68 m_{3}^{2}\right)
 t^{3}+\Big(360 m_{1}^{4}-780 m_{1}^{2} m_{2}^{2}-780 m_{1}^{2}
   m_{3}^{2}\nonumber\\&\quad +436 m_{2}^{4}-904 m_{2}^{2} m_{3}^{2}+436
   m_{3}^{4}\Big) t^{2}+\Big(-136 m_{1}^{6}+324 m_{1}^{4}
   m_{2}^{2}+324 m_{1}^{4} m_{3}^{2}-256 m_{1}^{2} m_{2}^{4}\nonumber\\&\quad -256
   m_{1}^{2} m_{3}^{4}+68 m_{2}^{6}-68 m_{2}^{4} m_{3}^{2}-68
   m_{2}^{2} m_{3}^{4}+68 m_{3}^{6}\Big) t -28   \prod_{i=1}^4\mu_i\nonumber\\&\quad \times\Big(2 m_{1}^{4}-m_{1}^{2} m_{2}^{2}-m_{1}^{2} m_{3}^{2}-m_{2}^{4}+2 m_{2}^{2} m_{3}^{2}-m_{3}^{4}\Big) \nonumber
\end{align}
with $c^{(1)}_2(m_1,m_2,m_3)=c^{(1)}_1(m_2,m_1,m_3)$, and $c^{(1)}_3(m_1,m_2,m_3)=c^{(1)}_1(m_3,m_2,m_1)$, 
and $c^{(1)}_1(\vec m)+c_2^{(1)}(\vec m)+c_3^{(1)}(\vec m)=4
\mathscr{S}_\su^0(\vec m,t)$.

      \subsection{The two-point one-mass kite}\label{sec:kite}
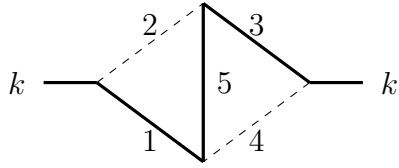
\begin{figure}[h]
	\centering\begin{tikzpicture}[scale=0.7]
		\draw [black, very thick] (-2,0) to (-3,0);
		\draw [black, very thick] (2,0) to (3,0);
		\draw [black, dashed] (-2,0) to (-0,1.5);
		\draw [black, very thick] (-2,0) to (-0,-1.5);
		\draw [black, very thick] (2,0) to (0,1.5);
		\draw [black, dashed] (2,0) to (0,-1.5);
		\draw[black, very thick] (0,1.5) to (0,-1.5);
		\node[text width=0.5cm, text centered ] at (-1.,1.1) {$2$};
		\node[text width=0.5cm, text centered ] at (-1.,-1.1) {$1$};
		\node[text width=0.5cm, text centered ] at (1.,1.1) {$3$};
		\node[text width=0.5cm, text centered ] at (1.,-1.1) {$4$};
		\node[text width=0.5cm, text centered ] at (0.4,0.) {$5$};
		\node[text width=0.5cm, text centered ] at (-3.5,0.0)
                {$k$};
                	\node[text width=0.5cm, text centered ] at (3.5,0.0) {$k$};
	\end{tikzpicture}
	\caption{Special two-point kite. Dashed lines are massless
          propagators and solid lines are massive propagators. The
          labels of the graph give the index of the edge variable $x_i$.}\label{fig:kite1mass}
\end{figure}
We consider the two points  kite graph of Fig.~\ref{fig:kite1mass} with three massive and
          two massless propagators. Its Symanzik polynomials read
\begin{equation}
	\textbf U_{\Kite} =(x_1+x_2) (x_3+x_4)+(x_1+x_2+x_3+x_4) x_5,
\end{equation}
\begin{multline}
\textbf F _{\Kite}(m^2,p^2)=k^2 (x_1 x_2 x_3 x_4\sum_{i=1}^4 x_i^{-1}+(x_1+x_4)(x_2+ x_3) x_5)
-m^2 (x_1+x_3+x_5) 	\textbf U_{\Kite} \,.
\end{multline}
Now setting $k^2=X m^2$ we have  single scale problem and now set $m=1$ so
\begin{equation}
  \Omega_{\Kite}^\epsilon=   {\textbf U_{\Kite}^{5-3\delta}\over
    \textbf F_{\Kite}(1,X)^{5-2\delta}}\left(\textbf U_{\Kite}^3\over
    \textbf F_{\Kite}^2\right)^\epsilon \Omega^{(5)}_0.
\end{equation}
It was shown
in~\cite{Broadhurst:1987ei,Adams:2016xah,Lairez:2022zkj} that the two point Kite graph with generic masses satisfies a
first order differential equation in four dimensions.
The integrand of the Feynman integral is the twisted differential form

\begin{equation}
\Omega_{\Kite} ={	\textbf U_{\Kite}^{5-3\delta} \over \textbf (F _{\Kite}(X))^{5-2\delta}}\,\left(	\textbf U_{\Kite}^3\over \textbf F _{\Kite}(X)^2\right)^\epsilon\,\Omega^{(5)}_0
\end{equation}
and the result of the reduction gives the  differential operator 
 \begin{equation}
	\label{eq:Kite}
	\mathscr{L}^\epsilon_{\Kite}=X(X-1){d\over dX} +X-1+(1+X)\epsilon\,.
\end{equation}

\subsection{The three-point ice-cream cone graph}\label{sec:ice-cream}

\begin{figure}[h]
	\centering
	\begin{tikzpicture}[scale=0.6]
		\filldraw [color = black, fill=none, very thick] (0,0) circle (2cm);
	\draw [black,very thick] (-2,0) to (2,0);
	\draw [black,very thick] (-2,0) to (-3,0);
	\draw [black,very thick] (2,0) to (3,0);
	\draw [black,very thick] (0,-2) to (0,-3);
	\node[text width=0.5cm, text centered ] at (-1.2,-1.) {$y_1$};
	\node[text width=0.5cm, text centered ] at (0,1.5) {$x_1$};
	\node[text width=0.5cm, text centered ] at (1.3,-1.) {$y_2$};
	\node[text width=0.5cm, text centered ] at (0.,0.25) {$z$};
	\node[text width=0.5cm, text centered ] at (-3.5,0.0) {$k_1$};
	\node[text width=0.5cm, text centered ] at (0,-3.5) {$k_2$};
	\node[text width=0.5cm, text centered ] at (3.5,0) {$k_3$};
	\end{tikzpicture}
	\caption{The ice cream cone graph. The  massive external momenta are
          $k_i$ satisfy $k_1+k_2+k_3=0$. We have labelled the graph with
          the edges variables.}\label{fig:icecream}
\end{figure}
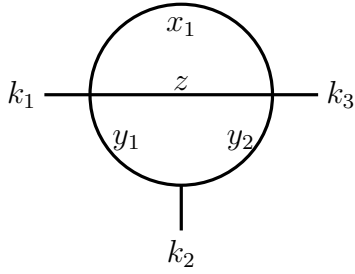

In this Section we give the result for the $\epsilon$-deformed
differential equation for the ice-cream cone graph of Fig.~\ref{fig:icecream} generalising the result
for $\epsilon=0$ given in~\cite{Lairez:2022zkj,Doran:2023yzu}.
The two-loop (one scoop) ice-cream cone  differential form in
$D=2-2\epsilon$ dimensions is given by
\begin{equation}
	\Omega_{\IceCream}^\epsilon(t)= {\textbf{U}_{\IceCream}\over
		\textbf{F}_{\IceCream}^2} \left(\textbf{U}_{\IceCream}^3\over  \textbf{F}_{\IceCream}^2\right)^{\epsilon}\,\Omega_0^{(4)},
\end{equation}

with
\begin{align}
	\textbf{U}_{\IceCream}&:=(y_1+y_2)(x_1+z)+zx_1,\cr
	\textbf{V}_{\IceCream}&:=k_2^2y_1y_2(z+x_1)+zx_1(k_1^2y_1+k_3^2y_2),\\
	\nonumber  \textbf{F}_{\IceCream}(t)&:=
	(\mu_1^2y_1+\mu_2^2y_2^2+m_1^2x_1+m_2^2z) \textbf{U}_{\IceCream}-t \textbf{V}_{\IceCream}.
\end{align}
We find the following results (some numerical cases are accessible on
the {\tt
  SageMath} worksheet \href{https://nbviewer.org/github/pierrevanhove/TwistedGriffithsDwork/blob/main/Worksheets/IceCream-Epsilon.ipynb}{IceCream-Epsilon.ipynb}).
\begin{itemize}
	\item {\bf The equal-kinematics case
		$\mu_1=\mu_2=m_1=m_2=k_1^2=k_2^2=k_3^2=1$:} the differential operator
	has order 3 and
	reads
	\begin{equation}
		\mathscr{L}_{\IceCream}^{[7],\epsilon}=\sum_{r=0}^4 \epsilon^r\,\mathscr{L}_{\IceCream}^{[7],r}
	\end{equation}
	with
	\begin{align}
		\mathscr{L}_{\IceCream}^{[7],0}&= 2 t^{3} \left(t -1\right)
		\left(t -3\right) \left(t -4\right) \left(d\over dt\right)^{3}+2 t^{2} \left(t
		-2\right) \left(11 t^{2}-44 t +15\right) \left(d\over dt\right)^{2}\nonumber\\&+2 t^{2}
		\left(29 t^{2}-116 t +89\right) {d\over dt} +32 t^{2} \left(t
		-2\right)\,,\\
		\mathscr{L}_{\IceCream}^{[7],1}&=t^{3} \left(t -1\right) \left(t -3\right) \left(t -4\right) \left(t +1\right) \left(d\over dt\right)^{3}\nonumber\\&+t^{2} \left(10 t^{4}-37 t^{3}-26 t^{2}+95 t +18\right) \left(d\over dt\right)^{2}\nonumber\\&+t \left(24 t^{4}+5 t^{3}-242 t^{2}+53 t +48\right)  {d\over dt} +12 t^{4}+64 t^{3}-112 t^{2}-48 t -12,
		\end{align}
	\begin{align}
		\mathscr{L}_{\IceCream}^{[7],2}&=t^{2} \left(t +1\right) \left(5 t^{3}-22 t^{2}+5 t +24\right) \left(d\over dt\right)^{2}+t^{2} \left(28 t^{3}-23 t^{2}-130 t -71\right)  {d\over dt} \nonumber\\&+26 t^{4}+56 t^{3}-48 t^{2}-64 t -18,\\
		\mathscr{L}_{\IceCream}^{[7],3}&=4 t \left(2 t^{2}-4 t -3\right) \left(t +1\right)^{2}
		{d\over dt}+6 \left(3 t^{2}-1\right) \left(t +1\right)^{2},\\
		\mathscr{L}_{\IceCream}^{[7],4}&= 4 t \left(t +1\right)^{3}.
	\end{align}
	The $\epsilon^0$ term factorises as
	\begin{multline}
		L_{\IceCream}^{[7],0}=\left(
		\left(2 t^{6}-16 t^{5}+38 t^{4}-24 t^{3}\right) {{d\over dt}} +4 \left(2 t^{3}-12 t^{2}+19 t -6\right) t^{2}\right)\cr\circ
		\left(
		{{d\over dt}} +\frac{5 t^{3}-30 t^{2}+49 t -18}{\left(t -4\right) t \left(t -1\right) \left(t -3\right)}
		\right)\circ \left({{d\over dt}} +\frac{2 t -4}{\left(t -1\right) \left(t -3\right)}
		\right).
	\end{multline}
	The  rightmost operator is the minimal differential equation for the
	$\epsilon=0$ case~\cite{Lairez:2022zkj}
	\begin{equation}
		L_{\IceCream}^{[7]}= {d\over dt}+ {2(t - 2)\over (t - 1)(t - 3)  }.
	\end{equation}

	\item \textbf{ The equal-mass case
		$\mu_1=\mu_2=m_1=m_2=1$ and generic momenta $k_1^2\neq
                k_2^2\neq
		k_3^2\neq 1$:} the differential operator is of order 3
	reads
	\begin{equation}
		\mathscr{L}_{\IceCream}^{[41^3],\epsilon}=\sum_{r=0}^1
		\mathscr{L}_{\IceCream,3}^{[41^3],r} \epsilon^r+ \sum_{r=0}^2   \mathscr{L}_{\IceCream,2-r}^{[41^3],r} \epsilon^{2+r},
	\end{equation}
	where $ \mathscr{L}_{\IceCream,n}^{[41^3],r}$  is of order $n$. The
	$\epsilon^0$ term factorises as
	\begin{equation}
		\mathscr{L}_{\IceCream,3}^{[41^3],0}=\mathscr{L}_{a,1}^0 \circ \mathscr{L}_{\IceCream}^{[41^3],0} \,,      
	\end{equation}
	where $\mathscr{L}_{a,1}^0$ is a first order operator   and the second order
	differential operator  $\mathscr{L}_{\IceCream}^{[41^3] ,0}
	$ matches the mass specialisation of the differential
	operator derived algorithmically in Section~5.2
	of~\cite{Lairez:2022zkj} and using Hodge theory in
	Section~7.3 of~\cite{Doran:2023yzu}.
	The highest order coefficient factorises as
	\begin{equation}
		\mathscr{L}_{\IceCream}^{[41^3],\epsilon}\Big|_{(d/dt)^3}=t^3(tk_2^2-(m_1+m_2)^2)(tk_2^2-(m_1-m_2)^2) c_1(t) c_2(t)c^{[41^3]}_3(t,\epsilon)   
	\end{equation}
	with
	\begin{multline}
		c_1(t)=   k_{1}^2 k_{2}^2 k_{3}^2 t^2 +t\Big(m_{1}^2 \left(-k_{1}^2 k_{3}^2-k_{2}^2 k_{3}^2+k_{3}^4\right)+m_{2}^2 \left(k_{1}^4-k_{1}^2
		k_{2}^2-k_{1}^2 k_{3}^2\right)\cr+(m_{3}+m_{4})^2
		\left(-k_{1}^2 k_{2}^2+k_{2}^4-k_{2}^2
		k_{3}^2\right)\Big)\cr
		+m_{1}^4 k_{3}^2+m_{1}^2 m_{2}^2 \left(-k_{1}^2+k_{2}^2-k_{3}^2\right)+m_{1}^2 (m_{3}+m_{4})^2
		\left(k_{1}^2-k_{2}^2-k_{3}^2\right)\cr+m_{2}^4 k_{1}^2+m_{2}^2 (m_{3}+m_{4})^2
		\left(-k_{1}^2-k_{2}^2+k_{3}^2\right)+k_{2}^2 (m_{3}+m_{4})^4,
	\end{multline}
	and
	\begin{multline}
		c_2(t)=t^2 k_{1}^2 k_{2}^2 k_{3}^2+t\Big(m_{1}^2 \left(-k_{1}^2 k_{3}^2-k_{2}^2 k_{3}^2+k_{3}^4\right)+m_{2}^2 \left(k_{1}^4-k_{1}^2
		k_{2}^2-k_{1}^2 k_{3}^2\right)\cr+(m_{3}-m_{4})^2 \left(-k_{1}^2 k_{2}^2+k_{2}^4-k_{2}^2 k_{3}^2\right)\Big)\cr
		+m_{1}^4 k_{3}^2+m_{1}^2 m_{2}^2 \left(-k_{1}^2+k_{2}^2-k_{3}^2\right)+m_{1}^2 (m_{3}-m_{4})^2
		\left(k_{1}^2-k_{2}^2-k_{3}^2\right)+m_{2}^4 k_{1}^2\cr+m_{2}^2 (m_{3}-m_{4})^2
		\left(-k_{1}^2-k_{2}^2+k_{3}^2\right)+k_{2}^2 (m_{3}-m_{4})^4,
	\end{multline}
	and $c^{[41^3]}_3(t,\epsilon)$ a polynomial of  degree 5 in $t$ and 1 in
	$\epsilon$. We recognise the physical thresholds of the ice-cream
	cone graph given in Section~5.2 of~\cite{Lairez:2022zkj} (and given
	on this
	page~\href{https://nbviewer.org/github/pierrevanhove/PicardFuchs/blob/main/PF-icecream-2loop.ipynb}{PF-icecream-2loop.ipynb}). The
	$\epsilon$ deformation only affects the position of the apparent singularities. 
	\item \textbf{The equal-mass case for the scoop
		$m_1=m_2=1$ and generic masses $\mu_1\neq\mu_2\neq1$ and generic
		momenta  $k_1^2\neq k_2^2\neq
		k_3^2\neq 1$:}  the differential operator has order 3 and has the
	$\epsilon$ expansion
	\begin{equation}
		\mathscr{L}_{\IceCream}^{[21^5],\epsilon}=\sum_{r=0}^1
		\mathscr{L}_{\IceCream,3}^{[21^5],r} \epsilon^r+ \sum_{r=0}^2   \mathscr{L}_{\IceCream,2-r}^{[21^5],r} \epsilon^{2+r},
	\end{equation}
	where $ \mathscr{L}_{\IceCream,n}^{[21^5],r}$  is of order $n$. The
	$\epsilon^0$ term factorises as
	\begin{equation}
		\mathscr{L}_{\IceCream,3}^{[21^5],0}=\mathscr{L}_{a,1}^0 \circ \mathscr{L}_{\IceCream}^{[21^5] ,0      } \, ,
	\end{equation}
	where $\mathscr{L}_{a,1}^0$ is a first order operator and the second order
	differential operator  $\mathscr{L}_{\IceCream}^{[21^5] ,0}
	$ matches the mass specialisation of the differential
	operator derived algorithmically in Section~5.2
	of~\cite{Lairez:2022zkj} and using Hodge theory in
	Section~7.3 of~\cite{Doran:2023yzu}.
	The leading coefficient factorises as
	\begin{equation}
		\mathscr{L}_{\IceCream,4}^{[21^5],\epsilon}\Big|_{(d/dt)^4}=t^3
		(tk_2^2-(m_1+m_2)^2)(tk_2^2-(m_1-m_2)^2) c_1(t)
		c_2(t) c^{[21^5]}_3(t,\epsilon)   .
	\end{equation}
	Only the positions of the apparent singularities 
	 depend on $\epsilon$. They arise from the
	roots of the polynomial $c^{[21^5]}_3(t,\epsilon)   $ of
	degree 5 in $t$ and 1 in $\epsilon$.
	\item \textbf{Generic masses non-vanishing
		$m_1\neq m_2\neq\mu_1\neq\mu_2$ and generic
		momenta  $k_1^2\neq k_2^2\neq
		k_3^2\neq 1$:}  the differential operator is of order 4 and has the
	$\epsilon$ expansion
	\begin{equation}
		\mathscr{L}_{\IceCream}^{[1^7],\epsilon}=\sum_{r=0}^1
		\mathscr{L}_{\IceCream,4}^{[1^7],r} \epsilon^r+ \sum_{r=0}^2
		\mathscr{L}_{\IceCream,2-r}^{[1^7],r} \epsilon^{2+r}.
	\end{equation}
	The
	$\epsilon^0$ term factorises as
	\begin{equation}
		\mathscr{L}_{\IceCream,4}^{[1^7],0}=\mathscr{L}_{a,1}^0 \circ \mathscr{L}_{b,1}^0 \circ \mathscr{L}_{\IceCream}^{[1^7] ,0      } \, ,
	\end{equation}
	where $\mathscr{L}_{a,1}^0$ and  $\mathscr{L}_{b,1}^0$ are
	first order operators and  the second order
	differential operator  $\mathscr{L}_{\IceCream}^{[1^7] ,0}
	$ matches the mass specialisation of the differential
	operator derived algorithmically in Section~5.2
	of~\cite{Lairez:2022zkj} and using Hodge theory in
	Section~7.3 of~\cite{Doran:2023yzu}.
	The leading coefficient factorises as
	\begin{equation}
		\mathscr{L}_{\IceCream,4}^{[1^7],\epsilon}\Big|_{(d/dt)^4}=t^4
		(tk_2^2-(m_1+m_2)^2)(tk_2^2-(m_1-m_2)^2) c_1(t)
		c_2(t) c^{[1^7]}_3(t,\epsilon)   .
	\end{equation}
	The position of the non-apparent singularities are
	not affected by the $\epsilon$ deformation but the
	apparent depend on $\epsilon$. They arise from the
	roots of the polynomial $c^{[1^7]}_3(t,\epsilon) $ of
	degree 11 in $t$ and 2 in $\epsilon$.
      \end{itemize}
      \subsection{The three-point non-planar  triangle-box  graph}\label{sec:chicken}
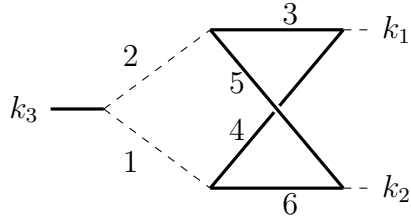
\begin{figure}[h]
	\centering	\begin{tikzpicture}[scale=0.7]
		 \draw [black, very thick] (-2,0) to (-3,0);
		\draw [black, dashed] (-2,0) to (-0,1.5);
		\draw [black, dashed] (-2,0) to (-0,-1.5);
		\draw [black,very thick] (0,1.5) to (2.5,-1.5);
		\draw [black,very thick] (0,-1.5) to (1.20,-0.06);
		\draw [black,very thick] (1.30,0.06) to (2.5,1.5);
		\draw [black,very thick] (0,1.5) to (2.5,1.5);
		\draw [black,very thick] (0,-1.5) to (2.5,-1.5);
                	\draw [black, dashed] (2.5,-1.5) to (3.0,-1.5);
		\draw [black, dashed] (2.5,1.5) to (3.0,1.5);
		\node[text width=0.5cm, text centered ] at (-1.5,-1.) {$1$};
		\node[text width=0.5cm, text centered ] at (-1.5,1.) {$2$};
		\node[text width=0.5cm, text centered ] at (1.5,1.8) {$3$};
		\node[text width=0.5cm, text centered ] at (1.5,-1.8) {$6$};
		\node[text width=0.5cm, text centered ] at (0.5,-0.4) {$4$};
		\node[text width=0.5cm, text centered ] at (0.5,0.5) {$5$};
		\node[text width=0.5cm, text centered ] at (-3.5,0.0) {$k_3$};
		\node[text width=0.5cm, text centered ] at (3.5,-1.5) {$k_2$};
		\node[text width=0.5cm, text centered ] at (3.5,1.5) {$k_1$};
	\end{tikzpicture}
	\caption{The non-planar  triangle-box   graph. Dashed lines are massless
          propagators and solid lines are massive propagators. The
          external momenta $k_i$ satisfy $k_1+k_2+k_3=0$ and $k_1^2=k_2^2=0$.}\label{fig:roastchicken}
\end{figure}
For the non-planar triangle-box  graph in Fig.~\ref{fig:roastchicken}, setting
the internal mass to 
$m=1$ and defining the  $2k_1 \cdot k_2 = X$ with $k_1^2=k_2^2=0$ and $k_1+k_2+k_3=0$. We have
\begin{equation}
	\textbf U_{\chickenDiag} = (x_1+x_2)(x_3+x_4+x_5+x_6)+(x_3+x_4)(x_5+x_6),
\end{equation}
\begin{multline}
\textbf F _{\chickenDiag}(X)=-\left((x_3+x_4+x_5+x_6)x_1x_2+x_1x_3x_5+x_2x_4x_6\right) X
\cr
+\left(x_3+x_4+x_5+x_6\right) 	\textbf U_{\chickenDiag} \, .
\end{multline}
The integrand of the Feynman integral is the twisted differential form
\begin{equation}
\Omega_{\chickenDiag} ={	\textbf U_{\chickenDiag}^{6-3\delta} \over \textbf (F _{\chickenDiag}(X))^{6-2\delta}}\,\left(	\textbf U_{\chickenDiag}^3\over \textbf F _{\chickenDiag}(X)^2\right)^\epsilon\,\Omega^{(6)}_0.
\end{equation}
an the differential operator is
 \begin{equation}
	\label{eq:roastchicken}
	\mathscr{L}^\epsilon_{\chickenDiag}=(16+X)X^2\left(d\over dX\right)^2+(2 (X+8) \epsilon +7 X+80) X{d\over dX} +4 (X+6) \epsilon +4 (2 X+15) \, .
      \end{equation}
\subsection{The four-point planar and non-planar double boxes graph}\label{sec:doubleboxes}
We show how to derive differential equation for  the massless box and massless double-box integrals in 
dimension $D=4-2\epsilon$. Unlike previous cases,  these integrals are divergent in four
dimensions so the $\epsilon=0$ integrals are not defined.

\subsubsection{The massless planar double-box graph}

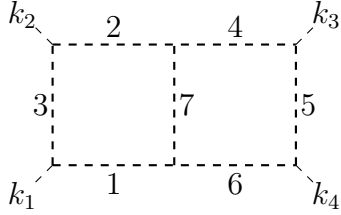
\begin{figure}[h]\centering
	\begin{tikzpicture}[scale=0.8]
		\draw[dashed](-1,-1)--(-1.3,-1.3);
		\draw[dashed](3,-1)--(3.3,-1.3);
		\draw[dashed](3,1)--(3.3,1.3);
		\draw[dashed](-1,1)--(-1.3,1.3);
		\draw[dashed,thick](-1,1)--(3,1);
		\draw[dashed,thick](-1,-1)--(3,-1);
		\draw[dashed,thick](-1,-1)--(-1,1);
		\draw[dashed,thick](1,-1)--(1,1);
		\draw[dashed,thick](3,-1)--(3,1);
		\node[text width=0.5cm, text centered ] at (-1.2,0) {$3$};
		\node[text width=0.5cm, text centered ] at (3.2,0) {$5$};
		\node[text width=0.5cm, text centered ] at (0,-1.3) {$1$};
		\node[text width=0.5cm, text centered ] at (2,-1.3) {$6$};
		\node[text width=0.5cm, text centered ] at (2,1.3) {$4$};
		\node[text width=0.5cm, text centered ] at (1.2, 0) {$7$};
		\node[text width=0.5cm, text centered ] at (0,1.3) {$2$};
		\node[text width=0.5cm, text centered ] at (-1.5,1.5){$k_2$};
		\node[text width=0.5cm, text centered ] at (3.5,1.5){$k_3$};
		\node[text width=0.5cm, text centered ] at (3.5,-1.5){$k_4$};
		\node[text width=0.5cm, text centered ] at (-1.5,-1.5){$k_1$};
	\end{tikzpicture} 
	\caption{The planar massless double-box graphs. The massless
          external momenta $k_i$ satisfy $k_1+\cdots +k_4=0$ and
          $k_i^2=0$. The labels of the graph give the index of the
          edge variable $x_i$. }\label{fig:doublebox}
	 \end{figure}
The graph polynomials associated to the massless double-box graph in
Fig.~\ref{fig:doublebox} are given by
\begin{align}
	\label{e:DoubleBoxgraphpolynomials}
	\textbf{U}_{\Box\!\Box}&=(x_1+x_2+x_3)(x_4+x_5+x_6)+(x_1+\cdots+x_6)x_7,\\
	\nonumber  \textbf{F}_{\Box\!\Box}(s,t)&=t x_3x_5x_7+s\bigg( (x_1+x_2+x_3) x_4x_6+(x_4+x_5+x_6)x_1x_2+(x_2+x_4)(x_1+x_6)x_7\bigg)
\end{align}
with the twisted differential in $D=4-2\epsilon$ in the projective
space $\mathbb P^6$
\begin{equation}\label{e:OmegaDoubleBox}
	\Omega_{\Box\!\Box}^\epsilon(s,t)=   {\textbf{U}_{\Box\!\Box}  \over
		\textbf{F}_{\Box\!\Box}(s,t)^{3}}\left(\textbf{U}_{\Box\!\Box}^3\over \textbf{F}^2_{\Box\!\Box}(s,t)\right)^\epsilon\,\Omega_0^{(7)} \, .
\end{equation}
We work with the single scale form $\tilde
\Omega_{\Box\!\Box}^\epsilon(X)=(-s)^{3+2\epsilon}
\Omega_{\Box\!\Box}^\epsilon(s,Xs)$
with the result
\begin{equation}\label{e:PF2box}
	\mathscr{L}_{\Box\!\Box}^\epsilon=(1+X)X^2 \left(d\over
	dX\right)^2+(2+3X+\epsilon) X{d\over dX}+X-\epsilon-2\epsilon^2.
\end{equation}

\subsubsection{The massless non-planar double-box graph}

\begin{figure}[h]\centering
 
	\begin{tikzpicture}[scale=0.6]
          \draw [black, dashed] (-2,1.5) to (0,1.5);
          		\draw [black, dashed] (-2,-1.5) to (0,-1.5);
		\draw [black,dashed] (-1,1.5) to (-1,-1.5);
		\draw [black,dashed] (0,1.5) to (2.5,-1.5);
		\draw [black,dashed] (0,-1.5) to (1.20,-0.06);
		\draw [black,dashed] (1.30,0.06) to (2.5,1.5);
                \draw [black,dashed] (0,1.5) to (2.5,1.5);
		\draw [black,dashed] (0,-1.5) to (2.5,-1.5);
		\draw [black,dashed] (2.5,-1.5) to (3.0,-1.5);
		\draw [black,dashed] (2.5,1.5) to (3.0,1.5);
                \node[text width=0.5cm, text centered ] at (-1.4,0) {$4$};
		\node[text width=0.5cm, text centered ] at (2.1,-0.5) {$7$};
		\node[text width=0.5cm, text centered ] at (-0.5,-2) {$3$};
		\node[text width=0.5cm, text centered ] at (1.5,-2) {$6$};
		\node[text width=0.5cm, text centered ] at (1.5,2) {$2$};
		\node[text width=0.5cm, text centered ] at (0.4, 0.5) {$5$};
		\node[text width=0.5cm, text centered ] at (-0.5,2) {$1$};
		\node[text width=0.5cm, text centered ] at (-2.5,1.5){$k_2$};
		\node[text width=0.5cm, text centered ] at (3.5,1.5){$k_3$};
		\node[text width=0.5cm, text centered ] at (3.5,-1.5){$k_4$};
		\node[text width=0.5cm, text centered ] at (-2.5,-1.5){$k_1$};
	\end{tikzpicture}
	\caption{The non-planar massless double-box graphs. The massless
          external momenta $k_i$ satisfy $k_1+\cdots +k_4=0$ and
          $k_i^2=0$. The labels of the graph give the index of the
          edge variable $x_i$. }\label{fig:npdoublebox}
\end{figure}
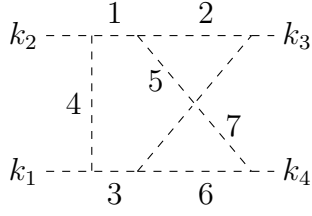

The graph polynomials associated with  the non-planar massless double-box
graph in Fig.~\ref{fig:npdoublebox} are given by
\begin{equation}
  \textbf{U}_{\DBoxNP}=(x_1+x_3+x_4)(x_2+x_5+x_6+x_7)+(x_2+x_7)(x_5+x_6) 
 \end{equation}
 and
 \begin{multline}
  \textbf{F}_{\DBoxNP}(s,t)=s\bigg(x_1x_3(x_2+x_5+x_6+x_7)+x_1x_6x_7+x_2(x_3x_5-x_4x_6)\bigg)\cr+t x_4 (x_5 x_7-x_2  x_6)
 \end{multline}
we work with the single scale differential form
 \begin{equation}
  \tilde \Omega^\epsilon_{\DBoxNP}=  (-s)^{3+2\epsilon}  {\textbf{U}_{\DBoxNP}  \over
		\textbf{F}_{\DBoxNP}(s,Xs)^{3}}\left(\textbf{U}_{\DBoxNP}^3\over \textbf{F}^2_{\DBoxNP}(s,sX)\right)^\epsilon\,\Omega_0^{(7)}\,.
 \end{equation}
The differential operator we obtain is
 \begin{multline}
   \mathscr{L}_{\DBoxNP}^\epsilon=(1+X)^2X^2\left(d\over dX\right)^2+(1+X)(1+2X)  (2+\epsilon) X{d\over dX} +2 X(X+1)
  \cr +\left(2 X(X+1)-1\right) \epsilon-2 \epsilon ^2 \,.
 \end{multline}

\subsection{The Witten ice-cream cone diagram}\label{sec:Wittenicecream}
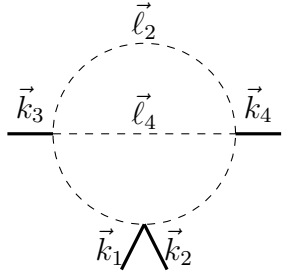
\begin{figure}\centering
\begin{tikzpicture}[scale=0.6]
\node at (0, 2.5) {$\vec \ell_2$};
\node at (0, 0.5) {$\vec \ell_4$};
\filldraw [color = black, fill=none, dashed] (0,0) circle (2cm);
\draw [ black,dashed] (-2,0) to (2,0);
\draw [ black,very thick] (-2,0) to (-3,0);
\draw [ black,very thick]  (3,0) to (2,0);
\draw [ black,very thick] (0,-2) to (0.5,-3);
\draw [ black,very thick] (0,-2) to (-0.5,-3);
\node at (-2.5, 0.65) {$\vec k_3$};
\node at (2.5, 0.65) {$\vec k_4$};
\node at (-0.75, -2.5) {$\vec k_1$};
\node at (0.75, -2.5) {$\vec k_2$};
\end{tikzpicture}
\caption{The Witten ice-cream cone graph in momentum space}\label{fig:Wittenicecream}
\end{figure}

We turn to the Witten ice-cream cone of Fig.~\ref{fig:Wittenicecream} in analytic regularisation
entering the two-loop correction to the cosmological correlator
between conformally coupled field analysed in Section~5.3.4 of~\cite{Chowdhury:2023arc}.
The cosmological correlator is the integration over the energy of the
two-loop flat space integral analytically regulated
\begin{equation}
  I_{\IceCream}= \int \frac{d^{4}L_{2}d^{4}L_{4}}{(L_2^2)^{1+\kappa}(L_4^2)^{1+\kappa}((L_4+L_2+Q)^2)^{1+\kappa}((L_4+L_2+\tilde Q)^2))^{1+\kappa} }  \,,
\end{equation}
where $\vec k_1+\vec k_2+\vec k_3+\vec k_4=0$ and $Q=(\omega_3,\vec
k_3)$ and $\tilde Q=(\omega_4,-\vec k_4)$.
The parametric representation is given by the integration 
$
  I_{\IceCream} = \int_{\Delta_4} \Omega_{\IceCream}^\kappa
$
over the domain $\Delta_4=\{x_i\geq0,1\leq i\leq 4\}$ of the
differential form
\begin{equation}
  \Omega_{\IceCream}^\kappa= {\pi^4\Gamma(4\kappa)\over
    \Gamma(1+\kappa)^4} \,{1\over \textbf{U}_{\IceCream}^{2} }
  \left(\prod_{i=1}^4 {x_i\textbf{U}_{\IceCream}\over \textbf{F}_{\IceCream}}  \right)^\kappa\,\Omega_0^{(4)}\,,
\end{equation}
with the graph polynomials
\begin{equation}
    \textbf{U}_{\IceCream}=x_1x_2+(x_1+x_2)(x_3+x_4),
  \end{equation}
  and
  \begin{equation}
    \textbf{F}_{\IceCream}=x_1x_2\left(x_3 Q^2+x_4 \tilde
      Q^2\right)+(x_1+x_2)x_3x_4 (Q-\tilde Q)^2.
  \end{equation}
  Setting  $u= Q^2/(Q-\tilde Q)^2$ and $v=\tilde Q^2/(Q-\tilde Q)^2$
  one finds the following Gr\"obner basis of differential operators\footnote{Details are provided in the  attached {\tt
     Mathematica}  notebook with  this example is accessible at \href{https://github.com/pierrevanhove/TwistedGriffithsDwork/blob/main/Mathematica/Icecream-AdS.pdf}{Icecream-AdS.pdf}.}
  \begin{align}
  \mathscr{L}_1&=\left(1-u
   -v\right) v\left(\partial\over\partial v\right)^{2} -2u
                 v{\partial\over\partial u}{\partial\over\partial
                 v}-(3 \kappa+1)u {\partial\over\partial u}+(3 \kappa
                 (1-  u-2  v)-v){\partial\over\partial v}-8 \kappa ^2,\\
  \mathscr{L}_2&=     -u\left(\partial\over\partial u\right)^{2}    +v\left(\partial\over\partial v\right)^{2}       -3 \kappa
   {\partial\over\partial u}+3 \kappa
                 {\partial\over\partial v},\\
 \mathscr{L}_3&=       2u^2 \left((u-v)^2+1-2(u+v)\right) \left(\partial\over
                \partial u\right)^3\\
                &+v\left(9 \kappa  \left((u+v-1)^2-2uv\right)-u^2+9 u^2-(v-1)^2-8u(1+v)\right)\left(\partial\over\partial
                  v\right)^2\cr
                  &+\left(\kappa^2\left(-29 u^2+u (43 v+29)-18 (v-1)^2\right)+ \kappa\left(6 u (v+3)-6 (v-1)^2\right)   +u (5 u-v+1)\right){\partial\over\partial u}\cr
                    &+\Big(\kappa ^2 \left(27 u^2-59 u v-54 u+36 v^2-63 v+27\right)-3 \kappa  \left(-9 u^2+6 u v+8 u+3 v^2-4 v+1\right)\cr
                      &-v (u+3 v-3)\Big){\partial\over\partial v}+24 (\kappa -1) \kappa ^2 (v-u-1)\,.\nonumber
  \end{align}
\section{Three- and higher-loop examples}\label{sec:higherloop}

\begin{figure}[h]
	\centering
	\begin{tikzpicture}[scale=0.6]
		\draw[very thick] (0,0) ellipse (3cm and 3cm);
		\draw[very thick] (0,0) ellipse (3cm and 2.cm);
		\draw [very thick](0,0) ellipse (3cm and 1.5cm);
		\draw [very thick](0,0) ellipse (3cm and 1.cm);
		\draw[very thick] (-3.5,0)--(3.5,0);
		\node[text width=0.5cm, text centered ] at (0,-.5) {$\vdots$};
		\node[text width=0.5cm, text centered ] at (0,0.5) {$\vdots$};
		\node [text width=0.5cm, text centered ] at (-4,0)
                {$p$};
                	\node [text width=0.5cm, text centered ] at (4,0) {$p$};
	\end{tikzpicture} 
	\caption{Multi-loop sunset}
  \label{fig:sunset}
\end{figure}
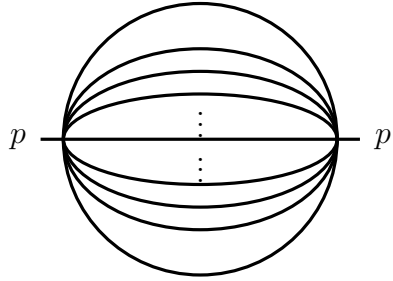

We now turn to the higher-loop sunset graph cases of Fig.~\ref{fig:sunset}. We first discuss the equal-mass case and then a numerical example at three-loop order with all
possible mass configurations.

\subsection{Minimal differential operator for higher-loop sunset}

We now consider the    $n-1$-loop sunset integral with $n\geq4$ in $D=2-2\epsilon$ dimensions. Eq.~\eqref{e:OmegaTwistDimReg} for this case reads
\begin{equation}
  I^{\epsilon}_{\su(n-1)}(\vec m,t,\epsilon)= \int_{\Delta_n} \Omega^{\epsilon}_{\su(n-1)}(\vec m,t,\epsilon)  ; \quad
  \Omega^\epsilon_{\su(n-1)}(\vec m, t,\epsilon)={\Omega^{(n)}_0\over
    \textbf{F}_{\su(n-1)}(t)}\left(\textbf{U}_{\su(n-1)}^n\over \textbf{F}_{\su(n-1)}(t)^{n-1}\right)^\epsilon
\end{equation}
with
\begin{align}
  \textbf{ U}_{\su(n-1)}&= x_1\cdots x_n\sum_{i=1}^n {1\over x_i}\,, \cr
      \textbf{   F}_{\su(n-1)}(t)&= \textbf{U}_{n-1} \sum_{i=1}^n m_i^2x_i-t x_1\cdots x_n \, .
\end{align}
Notice that $\textbf{U}_{\su(n-1)}^n/ \textbf{F}_{\su(n-1)}(t)^{n-1}$ is a homogeneous
rational function  of degree
0 in $(x_1,\dots,x_n)$. As usual the differential form is defined in
the complement of the vanishing locus of the denominator in $\mathbb P^{n-1}\backslash\{\textbf{F}_{\su(n-1)}(t)=0\}$.
In $D=2$ dimension ($\epsilon=0$) we have a rational  differential form
$ \Omega_{\su(n-1)}^{\epsilon}(\vec m,t,0)$. The differential operator has been
given up to six loops  for $\epsilon=0$ and it is in agreement with the Feynman integral
being a (relative) period of a Calabi--Yau  manifold of complex dimension $n-2$~\cite{Bloch:2013tra,Bloch:2014qca,Bloch:2016izu,Bourjaily:2019hmc,Bonisch:2020qmm,Bonisch:2021yfw,Candelas:2021lkc,Forum:2022lpz}.

   \subsubsection{The equal-mass case}\label{sec:highersunset1mass}
Already in $D=2$ dimensions, for the sunset integral from three loops on
 the Griffiths-Dwork algorithm had to be adapted because of
the non-isolated singularities of integrand. This was achieved
in~\cite{Lairez:2022zkj} by using syzygies. The resolution of the
linear system in Eq.~\eqref{e:sysCFU} also takes into account the syzygies
when including the $\epsilon$ dependent factor.

For the equal-mass case $m_1=\cdots =m_{l+1}=1$ we find the sunset
Feynman integral satisfies the differential equation
\begin{equation}
  \mathscr{L}_{\su(l)}^\epsilon I_\su(\{1,\dots,1\},t,\epsilon)=
  -(l+1)! {\Gamma(1+\epsilon)^l\over \Gamma(1+l\epsilon)}
\end{equation}
with
\begin{equation}
  \mathscr{L}_{\su(l)}^{\epsilon} =\sum_{r=0}^{l}  \mathscr{L}_{\su(l)}^{r} \epsilon^r \,,
\end{equation}
where the differential operator is $ \mathscr{L}_{\su(l)}^{r}$ is of
order $l-r$. In the all equal mass case the $\epsilon$
  dependence  takes the particular form
$
  \mathscr{L}_{\su(l)}^{\epsilon} = \mathscr{L}_{\su(l)}^{0}+ O(\epsilon)
  $
  where $\mathscr{L}_{\su(l)}^{0}$ is the differential operator of
  order $l$  for
  the all-equal-mass sunset integral in $D=2$ derived
  in~\cite{Vanhove:2014wqa} and the $\epsilon$ dependent differential
  operators have an order $l-r$ where $r$ is the power of $\epsilon$.
The order $\epsilon^0$ differential operator $ \mathscr{L}_{\su(l)}^{0}$ reproduces
the one derived in~\cite{Vanhove:2014wqa} up to five-loops using the
properties of the Feynman integral in $D=2$ dimensions (see as well~\cite{Bonisch:2020qmm,Pogel:2022yat,Pogel:2022ken,Pogel:2022vat,Mishnyakov:2023wpd,Mishnyakov:2023sly}).
By applying  the algorithm presented in
  section~\ref{sec:griff-dwork-reduct}  we derived the differential
  equations for the all-equal-mass sunset integral up to 20 loop
  orders.  The explicit results are given on the {\tt
  SageMath} worksheet
\href{https://nbviewer.org/github/pierrevanhove/TwistedGriffithsDwork/blob/main/Worksheets/Sunset-1mass-Epsilon.ipynb}{Sunset-1mass-Epsilon.ipynb}. We notice that the algorithm presented in this work produces 
the minimal  differential operator and does not need any factorisation
of the differential operator,  contrary to the procedure
presented in~\cite{Pogel:2022vat}. 


\subsubsection{The three-loop generic mass cases}\label{sec:threeloop}

For the three-loop sunset with different masses we find the following
results which are given as well  on the  {\tt
  SageMath} worksheet \href{https://nbviewer.org/github/pierrevanhove/TwistedGriffithsDwork/blob/main/Worksheets/Sunset-Threeloop-Epsilon.ipynb}{Sunset-Threeloop-Epsilon.ipynb}. With the notation of Section~4 of~\cite{Lairez:2022zkj}:
\begin{itemize}
\item  {\bf Case~$[4]$:} The equal-mass case $m_1=m_2=m_3=m_4$ has already been
  discussed in the previous section. The $\epsilon^0$ operator was derived and analysed in~\cite{Vanhove:2014wqa,Bloch:2014qca,Pogel:2022yat}. For $ \epsilon \ne 0$, the differential operator reads
   \end{itemize}
  \begin{multline}
    \mathscr{L}_{\su(3)}^{[4],\epsilon}=
    -(t - 16)  (t - 4)  t^2\left(d\over dt\right)^3 - 6  (t^3 -
                              15  t^2 + 32  t)  \left(d\over dt\right)^2 - (7  t^2 - 68  t +
                              64)  \left(d\over dt\right) - t + 4\cr
                              +\epsilon \left(-6  (t - 10  ) t^2 \left(d\over dt\right)^2 -
      6  (3  t - 20  ) t \left(d\over dt\right) +18- 6  t 
      \right)\cr
    +\epsilon^2\left(-(11  t^2 -
      28  t - 64)  \left(d\over dt\right) - 11  t + 14\right)+ \epsilon^3\left(-6  t - 12\right) \,.
  \end{multline}
\begin{itemize}
  \item   {\bf Case~$[31]$:} For two different masses $m_1=m_2=m_3 \neq m_4$ the
  differential operator is of order 5 and has the following $\epsilon$ dependence
  \begin{equation}
    \mathscr{L}^{[31],\epsilon}_{\su(3)}=       \sum_{r=0}^1 \epsilon^r
    \mathscr{L}^{[31],r}_{5}+  \sum_{r=0}^4 \epsilon^{2+r}   \mathscr{L}^{[31],r}_{4-r} \, ,
  \end{equation}
  where  $ \mathscr{L}^{[31],r}_{n}$ are of order $n$.
  The order $\epsilon^0$ operator factorises as
  \begin{equation}
         \mathscr{L}^{[31],0}_{5}=   \mathscr{L}^{[31],0}_{a,1} \circ \mathscr{L}^{[31],\epsilon}_{\su(3)} \, , 
       \end{equation}
        where  $ \mathscr{L}^{[31],0}_{a,1}$ is a  first order operator
       and $\mathscr{L}^{[31],0}_{\su(3)}$ is the fourth order
       differential operator of the
       three-loop sunset integral with mass configuration $[31]$ (see 
       in Section~4.3 of~\cite{Lairez:2022zkj}).
       The coefficient of the highest order term $(d/dt)^5$    is given by
       \begin{multline}
                   \mathscr{L}^{[31],\epsilon}_{\su(3)}\Big\vert_{(d/dt)^5}=
                   t^3  (t-(m_1-m_4)^2)(t-(m_1+m_4)^2)(t-(3m_1+m_4)^2) \cr\times (t-(3m_1-m_4)^2)
            q^{[31]}(t,\epsilon)      .
                 \end{multline}
                 The $\epsilon$ dependence appears only in the
                 apparent singularities determined by the polynomial
                 $q^{[31]}(t,\epsilon)$ of degree 3 in $t$ and 1 in $\epsilon$.
\item   {\bf Case~$[22]$:} For two different masses $m_1=m_2\neq m_3 = m_4$ the
  differential operator has order 6 and the $\epsilon$ expansion 
  \begin{equation}
    \mathscr{L}_{\su(3)}^{[22],\epsilon}=     \sum_{r=0}^3 \epsilon^r
    \mathscr{L}^{[22],r}_{6}+  \sum_{r=0}^5 \epsilon^{4+r}   \mathscr{L}^{[22],r}_{5-r} \, ,
  \end{equation}
  where  $ \mathscr{L}^{[22],r}_{n}$ are 
  operators of order $n$.
  The order $\epsilon^0$ operator factorises as
  \begin{equation}
         \mathscr{L}^{[22],0}_{6}=   \mathscr{L}^{[22],0}_{a,1} \circ \mathscr{L}^{[22],0}_{b,1} \circ\mathscr{L}^{[22],0}_{\su(3)} \,,
       \end{equation}
       where  $ \mathscr{L}^{[22],0}_{a,1}$ and $
       \mathscr{L}^{[22],0}_{b,1}$ are first order operators. 
        $\mathscr{L}^{[22],0}_{\su(3)}$ is the fourth order operator for the
       three-loop sunset integral with mass configuration $[22]$ given  Section~4.3 of~\cite{Lairez:2022zkj}.
    The coefficient of the highest order term $(d/dt)^6$    is given by
       \begin{multline}
                   \mathscr{L}^{[22],\epsilon}_{\su(3)}\Big\vert_{(d/dt)^6}=
                   t^4(t-(2m_1)^2)(t-(2m_4)^2)(t-(2m_1+2m_4)^2)\cr\times(t-(2m_1-2m_4)^2)
                   \, q^{[22]}(t,\epsilon).
                 \end{multline}
                 The $\epsilon$ dependence appears only in the
                 apparent singularities determined by the polynomial
                 $q^{[22]}(t,\epsilon)$ of degree 4 in $t$ and $3$ in $\epsilon$.
     \item   {\bf Case~$[211]$:} For three different masses $m_1=m_2\neq m_3 \neq m_4$ the
  differential operator has order 7 and has the $\epsilon$ expansion
  \begin{equation}
    \mathscr{L}^{[211],\epsilon}_{\su(3)}=       \sum_{r=0}^8 \epsilon^r
    \mathscr{L}^{[211],r}_{7}+  \sum_{r=0}^6 \epsilon^{9+r}   \mathscr{L}^{[22],r}_{6-r} \,,
  \end{equation}
   where  $ \mathscr{L}^{[211],r}_{n}$ 
  are operators of order $n$.
    The order $\epsilon^0$ operator factorises as
  \begin{equation}
         \mathscr{L}^{[211],0}_{7}=   \mathscr{L}^{[211],0}_{a,1} \circ  \mathscr{L}^{[211],0}_{b,1} \circ \mathscr{L}^{[211],0}_{c,1} \circ\mathscr{L}^{[211],0}_{\su(3)} \,,
       \end{equation}
        where  $ \mathscr{L}^{[211],0}_{a,1}$,  $
        \mathscr{L}^{[211],0}_{b,1}$ and  $ \mathscr{L}^{[211],0}_{c,1}$ are  first order operators
       and $\mathscr{L}^{[211],0}_{\su(3)}$ is the fifth order differential operator the
       three-loop sunset integral with mass configuration $[211]$
       given Section~4.3 of~\cite{Lairez:2022zkj}.   The coefficient of the highest order term $(d/dt)^6$    is given by
       \begin{multline}
                   \mathscr{L}^{[211],\epsilon}_{\su(3)}\Big\vert_{(d/dt)^6}=
                   t^5\left(t-(m_{1}-m_{2})^2\right) \left(t-(m_{1}+m_{2})^2\right)\cr\times
   \left(t-(m_{1}+m_{2}-2 m_{4})^2\right) \left(t-(m_{1}-m_{2}+2
   m_{4})^2\right) \left(t-(-m_{1}+m_{2}+2 m_{4})^2\right)\cr\times
   \left(t-(m_{1}+m_{2}+2 m_{4})^2\right)
                   \, q^{[211]}(t,\epsilon).
                 \end{multline}
                 The $\epsilon$ dependence appears only in the
                 apparent singularities determined by the polynomial
                 $q^{[211]}(t,\epsilon)$ of degree 9 in $t$  and 7 in $\epsilon$.
  \item   {\bf Case~$[1111]$:} For four different masses $m_1\neq m_2\neq m_3 \neq m_4$ the
 differential operator has order 11 and has the $\epsilon$ expansion 
  \begin{equation}
    \mathscr{L}_{\su(3)}^{[1111],\epsilon}=     \sum_{r=0}^{16}\epsilon^r
    \mathscr{L}^{[1111],r}_{11}+\sum_{r=0}^{11} \epsilon^{16+r}  \mathscr{L}^{[1111],\epsilon}_{11-r}\, .
  \end{equation}
    The order $\epsilon^0$ operator factorises as
  \begin{equation}
         \mathscr{L}^{[1111],0}_{0,11}=   \mathscr{L}^{[1111],0}_{a_1,1} \circ \cdots  \circ \mathscr{L}^{[1111],0}_{a_5,1}  \circ \mathscr{L}^{[1111],0}_{\su(3)}\,,
       \end{equation}
        where  $ \mathscr{L}^{[1111],\epsilon}_{a_1,1},\dots,  \mathscr{L}^{[1111],\epsilon}_{a_5,1}$ are  first order operators
       and $\mathscr{L}^{[1111],0}_{\su(3)}$ is the sixth  order differential operator for the
       three-loop sunset integral with mass configuration $[1111]$
       given in~\cite{Lairez:2022zkj}.
        The coefficient of the highest order term $(d/dt)^{11}$    is given by
       \begin{multline}
                   \mathscr{L}^{[1111],\epsilon}_{\su(3)}\Big\vert_{(d/dt)^{11}}=
                   t^{11}\left(t-(m_{1}+m_{2}-m_{3}-m_{4})^2\right) \cr\times
   \left(t-(m_{1}-m_{2}+m_{3}-m_{4})^2\right)
   \left(t-(m_{1}+m_{2}+m_{3}-m_{4})^2\right) \cr\times
   \left(t-(m_{1}-m_{2}-m_{3}+m_{4})^2\right)
   \left(t-(m_{1}+m_{2}-m_{3}+m_{4})^2\right)\cr\times
   \left(t-(m_{1}-m_{2}+m_{3}+m_{4})^2\right)
   \left(t-(-m_{1}+m_{2}+m_{3}+m_{4})^2\right) \cr\times
   \left(t-(m_{1}+m_{2}+m_{3}+m_{4})^2\right)
                   \, q^{[1111]}(t,\epsilon).
                 \end{multline}
                 The $\epsilon$ dependence appears only in the
                 apparent singularities determined by the polynomial
                 $q^{[1111]}(t,\epsilon)$ of degree 17 in
                 $t$ and 16 in $\epsilon$.
     \end{itemize}
\section{Summary and discussion}
In Section~\ref{sec:griff-dwork-reduct}, we have presented an
algorithm for deriving inhomogeneous differential equations satisfied by Feynman
integrals. At each derivative order the
 procedure consists of solving the linear systems~\eqref{e:sysCFU} in order to
determine the coefficients $c_{a_1,\dots,a_r}(\underline z)$ and the 
inhomogeneous term $\beta_\Gamma$ in~\eqref{e:PFOmegaGeneric}. Our
work introduces an explicit dependence on the regulators $\epsilon$ or $\kappa$, so it is worth discussing  their effect on the minimal differential
operator and on the singularities of the differential equations.
%

\subsection{Minimal order of the differential operator and number of
  master integrals}
\label{sec:minim-order-diff}
The minimal order differential operator gives direct
  information about the underlying algebraic-geometry associated with
  the Feynman integral. The knowledge of this operator is an essential
  ingredient for identifying if a given Feynman integral is a
  (relative) period associated with a genus 0, 1 or 2 curve,  a
  Calabi-Yau manifold or other object.

We start by summarising what we have found with the examples
  studied in this paper:

\begin{itemize}
	\item  For the $L$-loop
	sunset integrals, the number of irreducible master integrals 
	is $2^{L+1}-L-2$~\cite{Kalmykov:2012rr,Bitoun:2017nre,Kalmykov:2016lxx}.
	\begin{itemize}  \item The minimal differential operator for the all
		equal-mass case has order  the number of loops $L$.
		\item For generic mass configuration at one-loop we have an operator of order 1, at two-loop
	 an operator of order 4 and at three-loop  an operator
         has   order 11.
           \item In integer dimensions $\epsilon=0$ the order of the minimal order differential operator
  for generic kinematics can be  less than the number of master
  integrals.
 One typical
example is the one of the Picard-Fuchs operator
$\mathscr{L}_{\su(l)}$ for the multi-loop sunset  in $D=2$ dimensions which
has for  minimal order
$2^{L+1}-\binom{L+2}{\left\lfloor \frac{L+2}{2}\right\rfloor }$ for generic
mass configurations~\cite{Lairez:2022zkj}.  The order of the $\epsilon$-dependent
differential operator is the same as the  number of irreducible master integrals but its
$\epsilon$-independent piece $\mathscr{L}_{\su(l)}^0$ is factorisable as
$\mathscr{L}_{\su(l)}^0=\mathscr{L}_{1}\circ \mathscr{L}_{\su(l)}$.

	\end{itemize}
	\item For the two-loop ice-cream cone the number of irreducible
          masters is four.
          \begin{itemize}
            \item In the generic mass and kinematics case the minimal order of the $\epsilon$-deformed minimal differential
	operator is four.
        \item For special kinematic configurations  the ice-cream cone
          differential operator  is three which is smaller than the
          number of masters.
        \end{itemize}
        \end{itemize}

        We thus see that when $\epsilon\neq0$ and generic kinematics the order of the
		differential operator saturates the bound given by the number of
		irreducible masters.  This leads us to formulate the following observation:
       For general kinematics the minimal (i.e. not
factorisable) $\epsilon$-deformed differential operator has 
the same order
as the number of independent master integrals.
%

\subsection{Order reduction}

For special kinematics the minimal
  order is smaller than the number of master integrals.
  This reduction of order can be
understood by the factorisation of the differential operator
\begin{equation}
  \mathscr{L}_{\rm generic} \stackrel{\rm restriction}{\longrightarrow}
  \hat{\mathscr{L}}_{\rm restricted}   = \mathscr{L}_1 \circ\mathscr{L}_{\rm minimal} \,.
\end{equation}
The reduction of order arises when the integrand has more
singularities or more symmetries:
\begin{itemize}
  \item {\bf More singularities:} In the case of
massless internal lines or massless external kinematics, extra
singularities arise thus reducing the genus of the
singular locus of the integral.  This has for  consequence a lowering
of the order  minimal differential operator.

\item {\bf More symmetries:}
Another situation is when special choices of kinematic
configurations the integrand of the Feynman integral produce extra
symmetries in the space of projective variables $\underline
x$. This leads to new relations between independent (period) integrals, which reduce the number of independent integrals. A typical case is the
reduction order of the sunset integral according the mass
configurations as given
in~\cite{Bloch:2014qca,Lairez:2022zkj,Bonisch:2021yfw,Bonisch:2020qmm,Pogel:2022vat}. These
cases are not associated with the appearance of new singularities of
the Feynman integral.
\end{itemize}

In the case of special kinematic configurations that do not lead
to new singularities of the integral,  the order drop is not detected by  either the
computation of the Euler characteristic of complement of graph
hypersurface in the projective space of the edge variables, nor the
computation using the critical points of the
Lee-Pomeransky representation~\cite{Lee:2013hzt,Agostini:2022cgv,Fevola:2023kaw}   nor the
Baikov
representation~\cite{Frellesvig:2017aai,Frellesvig:2019uqt,Cacciatori:2021nli}.
For instance in~\cite{Marzucca:2023gto} it was explained how a hidden involution symmetry of the
two-loop non-planar double-box allows identifying the hyperelliptic
curve of genus 2 when the use of the Baikov representation gave a
curve of genus 3.
It is shown in~\cite{Doran:2023yzu}, using
a detailed  algebraic-geometrical analysis, that all planar two-loop
integrals are either period of rational curves, elliptic curve or genus
2 hyperelliptic curves or minimal order two or four respectively.

\subsection{The regulator dependence}
\label{sec:epsilon-dependence}

For a differential equation
\begin{equation}\label{e:PFconclusion}
c_N(z)  {d^Nf(z)\over dz^N}+\cdots + c_0(z) f(z)=0,
\end{equation}
the roots of $c_N(z)$ are the singularities of the differential
equation. A root of $c_N(z)$ where the solution $f(z)$ is
regular is called an apparent singularity. A root of $c_N(z)$
where the solution has a singularity is a real singularity (See 
Section~16.4 of~\cite{Ince} for details). 
For the case of Feynman integrals the non-apparent singularities
of~\eqref{e:PFconclusion} are the roots of the
discriminant of the singular locus of the integrand of  Feynman
integrals~\cite{Doran:2023yzu}.

We have noticed that the dimensional regulator $\epsilon$ appears only in  the
apparent singularities of the differential operator
$\mathscr{L}_\Gamma^\epsilon$. This means that the  $\epsilon$ deformation
does not change the position of the real singularities, but it affects
the local behaviour (the monodromy) of the solution near the
singularity.
The physical interpretation of this is that the kinematic
singularities (the position of the thresholds
and pseudo-thresholds) of a Feynman integral are independent of the
space-time dimension. However,  the local behaviour of the integral near
the thresholds and pseudo-thresholds does change with the space-time
dimension. The latter has been used with great success
when decomposing amplitudes using the generalised unitarity method~\cite{Bern:2011qt}.

\subsection{Outlook}\label{sec:conclusion}
We have presented a generalisation of the Griffiths-Dwork reduction
for deriving the differential operator acting on Feynman integrals in
dimensional regularisation or analytic regularisation. The algorithm
makes a special use of the fact that the twist from the regularisation
is the power of a degree zero homogeneous rational function in the
edge variables. The procedure amounts to
solving linear systems which is done using {\tt
  FiniteFlow} routines~\cite{Peraro:2019svx}.

We have applied the algorithm to various cases and derived the
inhomogeneous partial differential equations satisfied by the integrand of the Feynman integral in parametric representation.
In dimensional regularisation we have confirmed that the order the
differential operators is smaller or equal to the number of master
integrals. The order of the differential operators is lower for the cases of the kinematic invariants where the integrand
presents more symmetry (equal-masses, special kinematics,\dots) or
more singularities (massless internal or external states). Something
that was already noticed in the case of finite integrals~\cite{Lairez:2022zkj}  but stays
true for the regulated integrals.

One motivation for presenting this algorithm is its application to
Feynman integrals in analytical regularisation which arise in the evaluation of the cosmological
correlators~\cite{Heckelbacher:2022hbq,Chowdhury:2023khl,Chowdhury:2023arc},
since the commonly used  
integration-by-part algorithms need to be adapted to the case of
analytic regularisation with several propagators with generalised powers.

We have shown as well how to derive a Gr\"obner basis of partial
differential operators in some multiple scale cases. The differential
operators produced by the algorithm of this paper might arise as
specialisation of the system of partial differential operators
obtained by GKZ  approach. The restriction of the GKZ  D-module is a
difficult open problem, which we leave for further investigations.

\section*{Acknowledgements}
We would like to thank Pierre Lairez,  Eric Pichon-Pharabod  for discussions. We also thank Tiziano Peraro for correspondence on the use of {\tt
	FiniteFlow}. P.V. thanks the LAPTh for hospitality when this
      work has been completed.
The work of P.V. has received funding from the ANR grant ``SMAGP''
ANR-20-CE40-0026-01. P.V. acknowledges support of the Institut Henri
Poincar\'e (UAR 839 CNRS-Sorbonne Universit\'e), and LabEx CARMIN
(ANR-10-LABX-59-01). The work of LDLC is  supported by 
the European Research Council under grant ERC-AdG-885414.

\appendix\section{The Bessel representation for the sunset graphs}\label{sec:bessel}

Following the steps in Section~8 of~\cite{Vanhove:2014wqa}
gives the following Bessel integral representation for the multi-loop
sunset integrals in  $D=2-2\epsilon$ dimensions
\begin{multline}
  I_{\su(n-1)}^\epsilon(\vec m,t,\epsilon)=
  {2^{(n-1) (1-\epsilon)} t^{\epsilon\over2} \over    (m_1\cdots m_{L+1})^{\epsilon}
  \Gamma(1+(n-1)\epsilon)}\cr\times\int_0^\infty I_{-\epsilon}(\sqrt{t} x)
  \prod_{i=1}^{n}K_{-\epsilon}(m_ix)\,  x^{1+\epsilon(n-1)}dx.
\end{multline}
This integral representation is valid for  $t<(m_1+\cdots +m_{n})^2$.
For $x\to0$ we have
\begin{equation}
  \lim_{x\to0} I_{-\epsilon}(x)\simeq \left(x\over2\right)^{\epsilon};    \qquad
  \lim_{x\to0} K_{-\epsilon}(x)   \simeq \left(x\over2\right)^{\epsilon}
\end{equation}
and the integral converges as long as $2+(n-1)\epsilon>0$ which is the
condition $D=2-2\epsilon< 2n/(n-1)$ for the absence of ultraviolet divergences
for the $n-1$-loop massive sunset.

Using this representation and applying the creative telescoping
algorithm~\cite{Chyzak,Chyzak2,bostan2013creative,Koutchan} we have checked the results obtained the
extended Griffiths-Dwork reduction.
The creative telescoping algorithm builds an annihilator
\begin{equation}
  \mathscr{T}(t,\partial_t; x,\partial_x)= \mathscr{L}(t,\partial_t)+ \mathscr{C}(  t,\partial_t; x,\partial_x)
\end{equation}
of the integrand
$f(t,x):= I_{-\epsilon}(\sqrt{t} x)
  \prod_{i=1}^{L+1}K_{-\epsilon}(m_ix)\,  x^{1+\epsilon(n-1)}$
such that
\begin{equation}
      \mathscr{T}(t,\partial_t; x,\partial_x) f(t,x)=\mathscr{L}(t,\partial_t)i(t,x)+ \mathscr{C}(  t,\partial_t; x,\partial_x) f(t,x)=0
    \end{equation}
    implying that $\mathscr{L}(t,\partial_t)f(t,x)$ is the operator acting
    on the integral because the domain of integration is independent
    of $t$.
On an ordinary laptop, the results for the equal-mass case and up to twenty loops order, are obtained in
a few second to a few minutes, which is of the same order of time as
the generalised Griffiths-Dwork reduction presented in the main text
of this work.


\end{document}